\documentclass[12pt,a4paper]{article}
\usepackage{epsfig,amsfonts,amssymb}
\newcommand{\be}{\begin{equation}}
\newcommand{\ee}{\end{equation}}

\newcommand{\beq}{\begin{eqnarray}}
\newcommand{\eeq}{\end{eqnarray}}
\def\bgrk#1{\mbox{{\boldmath $#1$ \unboldmath}}\!\!}
\def\eq#1{(\ref{#1})}
\def\H1{\widehat{H}_1}

\newcommand{\pd}{\partial}

\begin{document}

\raggedbottom

\title{Semiclassical theory of spin-orbit interaction in the extended phase
      space}
\author{Mikhail Pletyukhov\thanks{E-mail:
mikhail.pletyukhov@physik.uni-regensburg.de} \ and  
Oleg Zaitsev\thanks{E-mail: oleg.zaitsev@physik.uni-regensburg.de}} 
\date{\emph{Institut f\"ur Theoretische Physik, Universit\"at Regensburg,\\
D-93040 Regensburg, Germany}}
\maketitle
\abstract{We consider the semiclassical theory in a joint phase space of spin
and orbital degrees of freedom. The method is developed from the path integral
using the spin-coherent-state representation, and yields the trace formula for
the density of states. We discuss the limits of weak and strong spin-orbit
coupling and relate the present theory to the earlier approaches.}
\vspace*{1cm}

\section{Introduction}

The subject of spin-orbit interaction in solid state physics has been
attracting much attention recently due to potential applications in spin-based
electronic devices \cite{sptr,tran}. It has often proved advantageous, on the
other hand, to exploit the semiclassical description of electrons in mesoscopic
systems in the ballistic regime \cite{riul}. Hence there is a substantial
interest in extending the semiclassical theory to include spin. Several
attempts were made in the last decade in this direction. Littlejohn and Flynn
\cite{lifl} revised and improved the asymptotic theory of coupled wave
equations and applied it to systems with standard spin-orbit coupling. Their
approach, however, relies on a finite strength of the spin-orbit interaction,
and thereby becomes invalid at the points in classical phase space where the
interaction vanishes---the so-called mode-conversion points. Frisk and Guhr
\cite{frgu} found that in certain cases this problem can be corrected by a
heuristic procedure. Bolte and Keppeler \cite{boke} studied the opposite
situation when the interaction is weak. They derived the trace formula for the
Dirac equation and Pauli equation with spin $1/2$. Based on the assumption that
the orbital motion is not affected by spin, their approach was used to explain
the anomalous magneto-oscillations in quasi-two-dimensional systems
\cite{wink}.  A comparative analysis of the above-mentioned semiclassical
methods and their applications to specific systems can be found in~\cite{cham}.

In a new semiclassical approach to spin-orbit coupling \cite{plet} the orbital
and spin degrees of freedom are combined in a joint (extended) phase space.
Then the whole arsenal of the semiclassical methods developed for spinless
systems can be applied there. The description of spin by continuous variables
is achieved by using the basis of coherent states. After a path integral
containing the spin and orbital coordinates is constructed, it can be evaluated
in the stationary-phase approximation, yielding the semiclassical propagator or
its trace.

A detailed investigation of the method of \cite{plet}, with the focus on the
trace formula for the density of states, is the subject of the present paper.
We will argue that the earlier approaches can be put under one roof by this new
theory. (A step in this direction was made in \cite{zait}.) Our discussion will
be limited to Hamiltonians linear in spin. For such systems, as we will show,
the semiclassical treatment is valid not only in the large-spin limit, but also
for a finite spin, including spin $1/2$. 

We expect that the semiclassical method presented below can be effectively used
in systems with relatively strong spin-orbit interaction, such as $p$-InAs or
InGaAs-InAlAs heterostructures \cite{inas}, or atomic nuclei \cite{nucl}.
Another field of application is molecular dynamics, where similar methods were
developed to map the discrete electronic states to continuous variables and
then to treat those semiclassically~\cite{thst}. 

Our paper is organized as follows. In the introductory section \ref{tfws} we
briefly review the derivation of the trace formula without spin, starting from
the path-integral representation for the propagator. Recently proposed by
Sugita \cite{sugi}, this new procedure reproducing the well-known result of
Gutzwiller \cite{gutz} will later be generalized to include spin. In Sec.\
\ref{scsqpf} we define the spin coherent states and write the
quantum-mechanical partition function in the path-integral representation for a
system with spin-orbit coupling. Sec.\ \ref{ssoc} is the main part of this
work, where three versions of the trace formula are derived. In the first two
instances we integrate over the spin variables in the path integral exactly and
then apply the stationary-phase approximation. Consequently, for weak
spin-orbit interaction (Sec.\ \ref{WCL} and Appendix \ref{ExQM}) we obtain the
generalization of the Bolte-Keppeler trace formula to arbitrary spin, and for
strong coupling (Sec.\ \ref{AL}) we recover, on a restricted basis, the result
by Littlejohn and Flynn (at the moment we cannot recover their ``no-name''
phase). In the third case (Sec.\ \ref{seps}) the path integral is evaluated by
the stationary-phase method in both spin and orbital variables. Then, using
Sugita's approach, we express the oscillating part of the density of states as
a sum over the periodic orbits in the extended phase space. Special attention
is paid to the Solari-Kochetov phase correction \cite{sola,koch,visa} arising
in the semiclassical limit of spin-coherent-state path integrals. In Sec.\
\ref{lcac} we classify the regimes of weak, strong, and intermediate coupling,
and show that the trace formula in the extended phase space works in all of
them. (Appendix \ref{adext} illustrates explicitly the strong-coupling limit.)
We also discuss and relax the large-spin applicability condition for the
semiclassical approximation. As an application of the general theory, in Sec.\
\ref{asm} we study two systems with Rashba-type spin-orbit interaction
\cite{rash} and the Jaynes-Cummings model~\cite{jacu}.

\section{Trace formula without spin}
\label{tfws}

In this section we summarize some basic facts related to the semiclassical
trace formula and present its derivation for systems without spin \cite{sugi}.
We consider a system described by the Schr\"odinger equation
\be
  \widehat H\, \psi_k = E_k \,\psi_k\,
\ee
with a discrete energy spectrum $\{E_k\}$. Its density of states $g(E) = \sum_k
\delta(E-E_k)$ can be subdivided into smooth and oscillating parts, i.e.,
\be
  g(E)= \widetilde{g}(E)+\delta g(E)\,.
\label{split}
\ee
From the semiclassical point of view, the smooth part $\widetilde g(E)$ is
given by the  contribution of all orbits with zero length and can be evaluated
by the (extended) Thomas-Fermi theory \cite{berm}. Numerically it can be
extracted by a Strutinsky averaging of the quantum spectrum \cite{book}. 

In this paper we assume the smooth part to be known and will be interested in
the  oscillating part $\delta  g(E)$, which is semiclassically approximated by
the trace formula \cite{gutz}
\be
 \delta g_{sc}(E) = \sum_{po} \mathcal A_{po}(E) 
                    \cos \left(\frac{1}{\hbar} \mathcal S_{po}(E) 
                    - \frac{\pi}{2} \sigma_{po} \right).
\label{trform}
\ee
The sum here is over all classical periodic orbits ($po$), including all
repetitions of each primitive periodic orbit ($ppo$). $\mathcal S_{po}(E)$ is
the action integral and $\sigma_{po}$ is the Maslov index of a periodic orbit.
The amplitude $\mathcal A_{po}(E)$ depends on the  integrability and the
continuous symmetries of the system. When all  periodic orbits are isolated in
phase space, the amplitude is given by 
\cite{gutz}
\be
  {\cal A}_{po}(E)=\frac 1 {\pi\hbar} \frac{T_{ppo}} 
                 {\sqrt{|\det ( \widetilde M_{po} - I_{2 (d - 1)})|}},
\label{amppo}
\ee
where $T_{ppo}$ is the period of the primitive orbit,   $ \widetilde M_{po}$ is
the stability matrix of the periodic orbit, and $d$ is the number of degrees of
freedom. $I$ is the unit matrix, the subscript denotes the dimensionality of
the space where $I$ acts. For two-dimensional systems $\sqrt{|\det (\widetilde
M_{po}- I_2)|} = 2 | \sin (\Lambda_{po} /2) |$, where $\Lambda_{po}$ is called 
the stability angle of a periodic orbit.

Recently Sugita \cite{sugi} has proposed a re-derivation of Gutz\-willer's
trace formula directly from the quantum partition function, avoiding the
calculation of the semiclassical propagator. Since we utilize his approach in
the following sections for systems with spin, we will briefly review it here.
Sugita's starting point is the quantum partition function (or the trace of the
quantum propagator) in the path integral representation
\be
  Z (T) = \int \! \frac {{\cal D} [{\bf q}] {\cal D} [{\bf p}]} {(2\pi
  \hbar)^d} \, \exp\{i{\cal R}[{\bf q,p};T]/\hbar\}.
\label{parti0}
\ee
The path integral is calculated along closed paths $(\mathbf q (t), \mathbf p
(t))$ in the $2 d$-dimensional phase space over a time interval $T$:
\be
  \frac {{\cal D} [{\bf q}] {\cal D} [{\bf p}]} {(2\pi \hbar)^d} 
  \equiv \lim_{N \to \infty} \prod_{l=1}^{N}
  \prod_{j=1}^d \left(\frac{d q_{j}(t_l) d p_{j}(t_l)} {2 \pi \hbar} \right) ,
  \quad t_l = l T/N. 
\ee
In \eq{parti0}
\be
  {\cal R} [{\bf q,p};T] \, =  \oint_0^T \biggl[ \frac12
         ({\bf p}\!\cdot\!{\bf\dot q} - {\bf q}\!\cdot\!{\bf\dot p})
         - \, {\cal H}({\bf q,p}) \biggr] dt 
\label{princ0}
\ee
is the Hamilton principal action function.  ${\cal H} ({\bf q,p})$ is the
classical Hamiltonian, i.e., the Wigner-Weyl symbol of $\widehat H$. The action
form ${\bf p} \cdot d {\bf q}$ can be antisymmetrized along the closed paths. 
The Fourier-Laplace transform of $Z (T)$ yields, after taking the imaginary
part, the density of states
\be
  g(E) =- \frac 1 \pi \,\mathrm{Im} \left( \frac{1}{i\hbar} \int_0^\infty
  e^{iET/\hbar}\,Z(T)\,dT\, \right).
\label{gofE}
\ee

The path integral \eq{parti0} receives its largest contributions from
the neighborhoods of the classical paths, along which the principal
function ${\cal R}$ is stationary according to Hamilton's variational
principle $\delta {\cal R} = 0$. The first variation hereby yields the
classical equations of motion
\be
  \dot{q}_i = \frac{\partial {\cal H}}{\partial p_i} , \quad
  \dot{p}_i = -\frac{\partial {\cal H}}{\partial q_i} .
\label{eom0}
\ee
One may evaluate the integrations in \eq{parti0} using the stationary-phase
approximation, which becomes asymptotically exact in the classical limit ${\cal
R}\gg\hbar$. The semiclassical approximation of the partition function $Z(T)$
then turns into a sum over all classical periodic orbits with fixed period $T$ 
\be
  {\cal Z}_{sc}(T) = \sum_{po} e^{i{\cal R}_{po}/\hbar}\!
                      \int \! \frac {{\cal D}[\bgrk{\eta}]} {(2\pi)^d}\,
                      \exp{\{i{\cal R}_{po}^{(2)}[\bgrk{\eta};T]\}}\,,
\label{zscl0}
\ee
where ${\cal R}_{po}$ are the principal functions \eq{princ0}, evaluated 
now along the classical orbits. The functional of the second variations is
\be
  {\cal R}_{po}^{(2)}[\bgrk{\eta},T] = \oint_0^T \left[\frac12\bgrk{\eta} 
          \cdot {\cal J} \dot{\bgrk{\eta}}-{\cal H}^{(2)}\right]\!dt
\label{Rpo2}
\ee
where $\bgrk{\eta}$ is the $2 d$-dimensional phase-space vector of small
variations $\bgrk{\eta} = (\bgrk{\lambda},\bgrk{\rho}) = (\delta{\bf q},
\delta{\bf p}) / \sqrt{\hbar}$ and {\small ${\cal J}=\left(\begin{array}{cc} 0
& I_d\! \\ \!\!-I_d & 0  \end{array}\right)$} is the $2 d$-dimensional unit
symplectic matrix. ${\cal H}^{(2)}$ is the second variation of the classical
Hamiltonian ${\cal H}$, calculated along the periodic orbits:
\be
  {\cal H}^{(2)} =  \frac{1}{2} \sum_{i,j = 1}^{d} \left[\lambda_i \lambda_j
  \frac{\partial^2{\cal H}}{\partial q_i \partial q_j} + 2 \lambda_i \rho_j
  \frac{\partial^2{\cal H}}{\partial q_i \partial p_j} + \rho_i \rho_j
  \frac{\partial^2{\cal H}}{\partial p_i \partial p_j}\right] .
\label{H2orb}
\ee   
Note that ${\cal Z}_{sc}(T)$ does not include the contribution of the
zero-length orbits.  

After a stationary-phase evaluation of the Fourier-Laplace integral \eq{gofE}
with ${\cal Z}_{sc}(T)$ instead of $Z (T)$, one finally obtains Gutzwiller's
trace formula \eq{trform} where the actions  ${\cal S}_{po}(E)={\cal
R}_{po}$+$ET_{po}$ are calculated at fixed energy $E$ and the periods of the
orbits are $T_{po}=dS_{po}/dE$.  The monodromy matrix $M_{po}$ is defined by 
$\bgrk{\eta}(T_{po})= M_{po}\, \bgrk{\eta}(0)$ in terms of the  solutions of
the linearized equations of motion $\dot{\bgrk{\eta}}={\cal J}\,\partial{\cal
H}^{(2)}\!/ \partial{\bgrk{\eta}}$, which are purely classical. After removing
the trivial parabolic block from $ M_{po}$ that appears due to the time
translation symmetry, one obtains the reduced  $2 (d -1)$-dimensional 
monodromy matrix $ \widetilde{M}_{po}$. The latter enters formula \eq{amppo}
and since it contains information about the stability of periodic orbits
\cite{book} it is often referred to as the stability matrix. For the Maslov
indices $\sigma_{po}$ Sugita has also given general formulae \cite{sugi}.

\section{Spin coherent states and quantum partition function}
\label{scsqpf}

We intend to generalize the path-integral representation for the partition
function \eq{parti0} to include the spin degree of freedom. The main issue then
is to be able to describe spin on the quantum-mechanical level by a continuous
variable. One can achieve this by using the overcomplete basis of spin coherent
states. The path integral for a system with spin in the SU(2)
spin-coherent-state representation originally appeared in a paper by Klauder
\cite{klau} as an integral on the sphere $\mathbb S^2$. Kuratsuji {\em et al.}\
\cite{kura} have represented it as an integral over paths in the extended
complex plane $\bar{\mathbb C}^1$. 

The SU(2) coherent state $|z;s\rangle$ for spin $s$ is defined by
\beq
  &~& |z; s \rangle = (1 + |z|^2)^{-s} \exp(z\hat{s}_{+})\, |s,m_s =
    -s\rangle\,,\\
  &~& \hat{s}_{-}|s, m_s = -s\rangle = 0\, ,
\eeq
where $z \in \bar{\mathbb C}^1$ is a complex number. The spin operators 
$\hat{s}_{\pm}=\hat{s}_1 \pm i\hat{s}_2$ and $\hat{s}_3$ are the
generators of the spin su(2) algebra:
\be
  [\hat{s}_3,\hat{s}_{\pm}] = \pm\hat{s}_{\pm}\,,\qquad
  [\hat{s}_{+},\hat{s}_{-}] = 2\hat{s}_3\,.\qquad
\ee
$|s,m_s \rangle$ are the eigenstates of $\hat{s}_3$. From the group-theoretical
point of view, $s \in \mathbb N/2$ labels irreducible representations of
SU(2).  

The irreducibility, as well as the existence of the SU(2) invariant  measure $d
\mu_s$, ensures that the resolution of unity holds in the spin-coherent-state
basis:
\beq
  \int \!|z;s\rangle\langle z;s| \,d\mu_s (z^* , z) = I_{2 s + 1}\,,\nonumber\\
  d \mu_s (z^* , z) = \frac{2s+1}{\pi\,(1+|z|^2)^2} d^{\,2}z\,.
\label{resun}
\eeq
This turns out to be the most important property of spin coherent states that
allows for the path-integral construction. The measure $d\mu_s$ takes account
of the curvature of the sphere $\mathbb S^2$.  In what follows, we denote
$|z;s\rangle$ simply by $|z\rangle$.

Let us now consider a quantum Hamiltonian with spin-orbit interaction
\be
  \widehat{H} = \widehat{H}_0 + \widehat{H}_s = \widehat{H}_0 (\hat{{\bf
  q}},\hat{{\bf p}})+ \hbar\, \hat{{\bf s}} \cdot \widehat{{\bf C}}({\hat{\bf
  q}},\hat{{\bf p}})\,,
\label{qh}
\ee
where $\hat{\bf s}=(\hat{s}_1,\hat{s}_2,\hat{s}_3)$ and $\widehat{\bf
C}=(\widehat{C}_1,\widehat{C}_2,\widehat{C}_3)$ is a  vector function of the
coordinate and momentum operators $\hat{\bf q}$, $\hat{\bf p}$. Thereby we
assume the most general form of spin-orbit interaction linear in spin. We write
the expression for the respective quantum propagator in terms of a path
integral in both the orbital variables ${\bf q,p}$ and the spin-coherent-state
variables $z,z^*$. Imposing periodic boundary conditions on the propagation and
thus integrating over closed paths, we arrive at  the expression for the
partition function~[cf.~\eq{parti0}]:
\be
  Z (T) = \int \! \frac {{\cal D} [{\bf q}] {\cal D} [{\bf p}]}{(2\pi \hbar)^d}
  \: {\cal D} \mu_s [z^* , z] \,\exp\,\{i{\cal R}[{\bf q,p},z^* , z;T]/\hbar\}.
\label{parti}
\ee
The Hamilton principal action function now includes the symplectic 1-form due
to spin:
\be
  {\cal R} [{\bf q,p},z^* , z;T] = \oint_0^T \biggl[ \frac12
         ({\bf p}\!\cdot\!{\bf\dot q} - {\bf q}\!\cdot\!{\bf\dot p})
         + \hbar s\,\frac{z \dot z^* - z^* \dot z}{i (1\!+\!|z|^2)} 
      -  {\cal H}({\bf q,p},z^*,z)\biggr] dt\,,
\label{princ}
\ee
and the path integration in \eq{parti} is taken over the $2(d+1)$-dimensional
extended phase space:
\be
  \frac {{\cal D} [{\bf q}] {\cal D} [{\bf p}]} {(2\pi \hbar)^d} \: {\cal D}
  \mu_s [z^* ,z] =  
  \lim_{N \to \infty} \prod_{l=1}^{N} \prod_{j=1}^d \frac{d q_{j}(t_l) d
  p_{j}(t_l)}  {2 \pi \hbar} \, d \mu_s (z^*(t_l) ,z(t_l))\,, 
\label{measure}
\ee
where the time interval $T$ is divided into $N$ time steps $t_l=lT/N$. 

The c-number Hamiltonian ${\cal H}({\bf q,p},z^* ,z)$ appearing in the
integrand of \eq{princ} is
\be
  {\cal H}({\bf q,p},z^* ,z) = {\cal
  H}_0({\bf q,p}) + {\cal H}_s({\bf q,p},z^* ,z) = {\cal H}_0({\bf q,p}) +
  \hbar s {\bf n}(z^* ,z)\cdot {\bf C} ({\bf q}, {\bf p})\, ,
\label{chs}
\ee
where ${\cal H}_0({\bf q,p})$ and ${\bf C}({\bf q,p})$ are the Wigner-Weyl
symbols of the operators $\widehat{H}_0$ and $\widehat{{\bf C}}$, and ${\bf
n}=(n_1,n_2,n_3)=\langle z|\hat{\bf s}|z\rangle/s$ is the unit vector of
dimensionless classical spin. Its components in terms of $z^*$ and $z$ are
given by 
\be
  n_1 + i n_2= 2 z^* /(1+|z|^2), \quad n_3=-(1-|z|^2)/(1+|z|^2).
\label{clsp}
\ee
One may recognize the stereographic projection from a point ${\bf n}$ on the
unit Bloch sphere onto a point $(\mathrm{Re}\: z, - \mathrm{Im}\: z)$ on the
complex plane, whereby the south pole is mapped onto its origin $z=0$.

We remark that the choice of the c-number Hamiltonian is not unique and, in the
discretized path integral, is linked to a specific discretization prescription
\cite{lang}. For symbolic manipulations in the continuous limit, it is often
convenient to use the Wigner-Weyl symbol of the quantum Hamiltonian as the
c-number Hamiltonian. Note, however, that $s {\bf n}$ is the covariant symbol
of the spin operator $\hat {\mathbf s}$, and not the Wigner-Weyl symbol. The
consequences of the present choice of ${\cal H}({\bf q,p},z^* ,z)$ \eq{chs} for
the semiclassical theory are discussed in Sec.\ \ref{lcac}.

\section{Semiclassics with spin-orbit coupling}
\label{ssoc}

In this section we consider the semiclassical approach for systems with
spin-orbit interaction. We begin with the limit of weak coupling (Sec.\ 
\ref{WCL}), calculating the spin part of the path integral exactly. The trace
formula we obtain is a generalization of the results by Bolte and Keppeler
\cite{boke} to arbitrary spin. A similar approach leads, under certain
restrictions, to the trace formula in the adiabatic (strong-coupling) limit
(Sec.\ \ref{AL}), that was studied by Littlejohn and Flynn \cite{lifl}.  In
Sec.\ \ref{seps} we present a new method that treats both orbital and spin
degrees of freedom semiclassically \cite{plet}. With the latter approach one
can go beyond the limits of weak and strong coupling. The relationship between
the methods, in connection with the coupling strength, is analyzed in Sec.\
\ref{lcac}.

\subsection{Weak-coupling limit: quantum-mechanical treatment of spin}
\label{WCL}

We derive here the semiclassical trace formula for the Hamiltonian \eq{qh} in
the asymptotic limit $\hbar \rightarrow 0$. In addition to the standard
semiclassical requirement ${\cal R}/ \hbar  \gg 1$, which is automatically
fulfilled, this limit also implies that the spin-orbit interaction energy is
small:
\be 
  |{\cal H}_s({\bf q,p},z^* ,z)| \ll |{\cal
  H}_0({\bf q,p})|,
\label{wclcond}
\ee
i.e., the system is in the weak-coupling regime. 

It is convenient to separate the terms that explicitly contain $\hbar$ in the
Hamilton principal action function \eq{princ}. Representing 
\be 
  {\cal R} [{\bf q}, {\bf p}, z^* ,z;T] = {\cal R}_0 [{\bf q}, {\bf p} ; T]
  + \hbar s {\cal R}_1 [{\bf q}, {\bf p}, z^* ,z;T],
\label{Rpart}
\ee
where the unperturbed part ${\cal R}_0$ is given by \eq{princ0} with the
Hamiltonian ${\cal H}_0({\bf q,p})$ and 
\be
  {\cal R}_1 [{\bf q}, {\bf p}, z^* ,z;T] = \oint_0^T \biggl[\frac{z \dot
  z^* - z^* \dot z}{i (1\!+\!|z|^2)}  -  {\bf n}(z^* ,z)\cdot {\bf C} ({\bf q},
  {\bf p})\biggr] dt\,, 
\ee
we can write the partition function \eq{parti} in the form
\be
  Z (T) = \int \frac {{\cal D}[{\bf q}]{\cal D} [{\bf p}]} {(2\pi \hbar)^d}
  {\cal M}_s [{\bf q}, {\bf p};T] \exp \left\{ i {\cal R}_0 [{\bf q}, {\bf
  p};T]/ \hbar \right\}
\label{seppf}
\ee
with
\be
  {\cal M}_s [{\bf q}, {\bf p};T] = \int {\cal D} \mu_s [z^* , z] \exp
  \left\{ i s  {\cal R}_1 [{\bf q}, {\bf p}, z^* , z;T]  \right\}.
\label{Ms1}
\ee

Now we can evaluate $Z (T)$ in the stationary-phase approximation in ${\bf q}$
and ${\bf p}$. Since $s {\cal R}_1 \ll {\cal R}_0/ \hbar$ in the case under
consideration, ${\cal M}_s [{\bf q}, {\bf p};T]$ is a slowly varying functional
of ${\bf q}$ and ${\bf p}$ and need not be taken into account when evaluating
the stationary-phase condition. Applying the results of Sec.\ \ref{tfws}, we
obtain the semiclassical approximation to the partition function as a sum over
the \emph{unperturbed} periodic orbits (determined entirely by ${\cal H}_0$)
\be
  Z_{sc} (T) = \sum_{po} {\cal M}_{po} (T)\, {\cal A}^{(0)}_{po} (T)
  \exp \left\{ i {\cal R}^{(0)}_{po} /\hbar - \nu^{(0)}_{po} \pi /2 \right\}, 
\ee
where ${\cal A}^{(0)}_{po}$, $\;\nu^{(0)}_{po}$, ${\cal R}^{(0)}_{po} \equiv
{\cal R}_0 [{\bf q}^{(0)}_{po}, {\bf p}^{(0)}_{po};T]$, and the modulation
factor ${\cal M}_{po} (T) \equiv {\cal M}_s [{\bf q}^{(0)}_{po}, {\bf
p}^{(0)}_{po};T]$ are evaluated for the unperturbed trajectory. Performing the
Fourier-Laplace transform \eq{gofE} (also in stationary-phase approximation) we
arrive at the trace formula in the weak-coupling limit (WCL)
\be
  \delta g_{sc}^{\mathrm{WCL}} = \sum_{po} {\cal M}_{po} (E)\, {\cal
  A}^{(0)}_{po} (E) \cos \left\{ i {\cal S}^{(0)}_{po} /\hbar -
  \sigma^{(0)}_{po} \pi /2 \right\},
\label{trwclm}
\ee
which differs from the unperturbed trace formula for ${\cal H}_0$ only by the
presence of the modulation factor ${\cal M}_{po} (E) \equiv {\cal M}_{po}
(T^{(0)} (E))$ with $T^{(0)} (E) = d {\cal S}_{po}^{(0)} / dE$.

The modulation factor is determined in Appendix \ref{ExQM} by a direct
evaluation of the path integral \eq{Ms1}. For a Hamiltonian linear in spin, the
problem is effectively reduced to the calculation of the spin partition
function in the time-dependent external magnetic field ${\bf C} ({\bf q (t)},
{\bf p (t)}) \equiv (2 \,{\rm Re} f (t), - 2 \,{\rm Im} f (t), 2A (t))$,
determined by the path $({\bf q (t)}, {\bf p (t)})$. Then for a periodic orbit
one finds 
\be
  {\cal M}_{po} (E) = \frac{\sin [ (s + 1/2)\, \eta (E)]}{\sin [ \eta (E)/2]}
\label{modfs}
\ee
with 
\be
  \eta (E) = \int_0^{T^{(0)}} dt \, \{ 2 A^{(0)} (t) - [(f^{(0)} (t))^* z +
  f^{(0)} (t) z^*]\}.
\label{stangm}
\ee
$z^* (t)$ and $z (t)$ are to be found from the first-order differential
equation
\be
  \dot{z} = - i [ 2 A ^{(0)} (t) z + f^{(0)} (t) - (f^{(0)} (t))^* z^2 ] ,
  \quad z(0) = z(T),
\label{ztm} 
\ee
which is simply the precession equation $\dot{{\bf n}} = {\bf C}^{(0)} \times
{\bf n}$, ${\bf n} (0) = {\bf n} (T)$ for the classical spin vector ${\bf n}$
[Eq.\ \eq{clsp}]. This equation is angle-preserving, i.e., it rotates the Bloch
sphere without deforming it. Thus, the orientations of the Bloch sphere at
$t=0$ and $t=T$ are related by a rotation about some axis. The angle of
rotation is $\eta (E)$ (Appendix \ref{ExQM}). The choice of initial condition
${\bf n} (0)$ along this axis corresponds to a periodic solution (cf.\
\cite{zait}). Thus, \eq{ztm} has two periodic solutions, whose $\eta (E)$ have
opposite signs. Nevertheless, ${\cal M}_{po} (E)$ is well defined. Note that
$\eta (E)$ is equal to the part ${\cal R}_1 [{\bf q}^{(0)}_{po}, {\bf
p}^{(0)}_{po}, z^* , z;T^{(0)}]$ of the Hamilton principal action function
calculated along the periodic $z^* (t)$ and $z (t)$. Although we are able to
give a classical interpretation to the ingredients of ${\cal M}_{po} (E)$, no
stationary-phase approximation was used in its derivation. 

In the absence of spin-orbit interaction or external magnetic fied one finds 
${\cal M}_{po} (E) \! = 2s + 1$, i.e., the unperturbed trace formula is multiplied
by the spin degeneracy factor. 

For $s= 1/2$ the modulation factor ${\cal M}_{po} (E)$ was derived in
\cite{boke} by a different method, which, as ours, treats the spin degrees of
freedom on the quantum-mechanical level.  

Clearly, when the weak-coupling condition \eq{wclcond} breaks down, the trace
formula \eq{trwclm} is no longer valid. Indeed, in this case ${\cal R}_1$
is large, and ${\cal M}_s [{\bf q}, {\bf p};T]$ will influence the
stationary-phase condition for the integral \eq{seppf}.

Note that the representation of terms in the WCL trace formula as a product of
the unperturbed part and the modulation factor [Eq.\ \eq{trwclm}] remains
valid even when the Hamiltonian is nonlinear in spin. However, the simple
expression for the modulation factor Eq.\ \eq{modfs} is specific to the
Hamiltonian linear in spin.

\subsection{Adiabatic limit: strong coupling}
\label{AL}

With an exact integration over the spin degrees of freedom in the path integral
we can also derive the trace formula in the adiabatic limit. This limit is
defined by the requirement 
\be
  |{\bf C(q,p)}| \gg T^{-1}_{\mathrm{orb}},
\label{adiab}
\ee
where $T_{\mathrm{orb}}$ is the period of the orbital motion. Since $|{\bf C}|$
is the frequency of precession of the classical spin about the instantaneous
magnetic field ${\bf C}$, Eq.\ \eq{adiab}  means that the spin motion is much
faster than the orbital motion. The results of this section will be most useful
in the strong-coupling limit (SCL) $|{\cal H}_s| \sim |{\cal H}_0|$ or $|{\bf
C}| \sim |{\cal H}_0| /\hbar$, where \eq{adiab} is automatically satisfied.
Formally speaking, the SCL can be stated as the double-limit $\hbar \to 0$,
$|{\bf C}| \to \infty$ with $\hbar |{\bf C}| = \mathrm{const}$ (cf.\
\cite{boke}).

We start with the representation \eq{seppf} for the partition function and
write the prefactor in the form (Appendix \ref{ExQM})
\be
  {\cal M}_s [{\bf q}, {\bf p};T] = \frac{\sin \{ (s + 1/2)\, \eta\,
  [\mathbf{q,p};T]\}}{\sin \{ \eta \,[\mathbf{q,p};T]/2\}} = \sum_{m_s =-s}^{s}
  \exp \{i m_s \eta\, [\mathbf{q,p};T]\}.
\label{Msexact}
\ee
Then $Z (T)$ becomes a sum over polarizations
\be
  Z (T) = \sum_{m_s =-s}^{s} \int \frac {{\cal D}[{\bf q}]{\cal D} [{\bf p}]}
  {(2\pi \hbar)^d} \exp \left\{ i {\cal R}_0 [\mathbf{q,p};T]/ \hbar + i m_s
  \eta\, [\mathbf{q,p};T] \right\},
\label{ZAdLim}
\ee
where the path integrals will be calculated by the stationary phase. The
functional $\eta\, [\mathbf{q,p};T]$ is given by [cf.\ \eq{stangm}]
\be
  \eta \, [\mathbf{q,p};T]= \int_0^T dt \, \{ 2 A - [f^* z + f z^*]\}
\label{etaqpt}
\ee
with $f = |{\bf C}| \sin \theta_C e^{-i \phi_C}$ and $A= (|{\bf C}| /2) \cos
\theta_C$ in terms of the polar angles $\theta_C$, $\phi_C$ of the vector ${\bf
C}$. The trajectory $z(t)$ is one of the two periodic solutions of the equation 
\be
   \dot{{\bf n}} = {\bf C} \times {\bf n}, \quad |{\bf n}| = 1,
 \label{preceq}
\ee
which can be solved approximately in the adiabatic limit. We look for the
solution in the form ${\bf n} =  {\bf n}^{(0)} + {\bf n}^{(1)}$, where ${\bf
n}^{(0)} \parallel {\bf C}$ and ${\bf n}^{(1)} \perp {\bf C}$. Assuming ${\bf
n}^{(1)} \sim {\bf n}^{(0)}/ (|{\bf C}|\, T_{\mathrm{orb}}) \ll {\bf n}^{(0)}$,
we obtain 
\be
  {\bf n}^{(0)} = \pm \frac{{\bf C}}{|{\bf C}|}, \quad {\bf n}^{(1)} = \mp
  \frac{{\bf n}^{(0)} \times \dot{{\bf n}}^{(0)}}{| {\bf C}|}.
\ee 
Then \eq{etaqpt} gives us the phase
\be
  m_s\, \eta \, [\mathbf{q,p};T]= -m_s\,\int_0^T |{\bf C}|\, dt\, +\,
  \varphi_B^{(m_s)} \, +\, \mathcal O (|{\bf C}|\, T_{\mathrm{orb}})^{-1}.
\label{etaAdLim}
\ee
Here the second term is the Berry phase 
\be
  \varphi_B^{(m_s)} = m_s \int_0^T (1 + \cos \theta_C )\, \dot{\phi}_C\,  dt.
\label{Berry}
\ee

When the path integrals \eq{ZAdLim} for $Z(T)$ are evaluated by stationary
phase, the Berry phase $\varphi_B^{(m_s)} \sim 1$ does not play a role in the
determination of the stationary-phase point. In the SCL, i.e., when $|{\bf C}|
\sim |{\cal H}_0|/ \hbar$, the first term of \eq{etaAdLim} must be varied
together with ${\cal R}_0 / \hbar$ in order to derive the stationary-phase
condition. Then after the standard procedure (Sec.\ \ref{tfws}) we obtain the
trace formula 
\be
  \delta g_{sc}^{\mathrm{SCL}} = \sum_{m_s =-s}^{s} \delta
  g_{sc}^{(m_s)},
\label{trfSCL}
\ee
where each polarized density of states $\delta g_{sc}^{(m_s)}$ is given by
\eq{trform} with the classical dynamics controlled by the effective Hamiltonian
\be
  {\cal H}_{\mathrm{eff}}^{(m_s)} \mathbf{(q,p)}= {\cal H}_0 \mathbf{(q,p)} +
  m_s \hbar |\mathbf{C(q,p)}|
\label{Heff}
\ee
and the Berry phase added. An independent semiclassical derivation of this
result is given in Appendix \ref{adext}.

Actually, this result is valid only if ${\bf C(q,p)}$ depends on either $q_j$
or $p_j$, but not both, for each $j = 1, \ldots, d$. In the latter case such
symbolic manipulations with path integrals do not seem to be valid, and, as a
result, the ``no-name'' term found by Littlejohn and Flynn \cite{lifl} is
missing. The problem of recovering this term in the path-integral approach has
been known for some time and was already mentioned in Ref.~\cite{lifl}. Making
some heuristic assumptions, Fukui \cite{fuku}  reproduced the no-name term for
the Jaynes-Cummings model in a framework similar to ours. It is not clear to us
at this stage, whether his approach can be generalized to a wider class of
Hamiltonians.  We suppose that this term emerges due to the operator ordering,
and hope to recover it making a careful analysis of the continuous limit of the
path integral.

It may happen that the adiabatic condition \eq{adiab} is fulfilled in the WCL.
The trace formula is well defined if the limit $\hbar |{\bf C}| / {\cal H}_0
\rightarrow 0$ is taken \emph{before} the limit $|{\bf C}|\, T_{\mathrm{orb}}
\rightarrow \infty$. Hence, $m_s\, \eta \, [\mathbf{q,p};T]$ does not
contribute to the determination of the stationary point, and the orbital motion
is governed entirely by ${\cal H}_0$. The results of the previous section
will be recovered with $\eta (E)$ given by \eq{etaAdLim}, i.e., the fast spin
precession can be removed from $\eta (E)$ in the adiabatic limit. 

We note that expression \eq{etaAdLim} becomes invalid if the adiabatic
condition \eq{adiab} breaks down anywhere along the periodic orbit. In
particular, this occurs in the mode-conversion points defined by
$\mathbf{C(q,p)} = 0$.

\subsection{Semiclassics in the extended phase space}
\label{seps}

In the present approach, proposed recently in \cite{plet}, both the orbital and
spin degrees of freedom are treated semiclassically. For the formal  derivation
we assume the standard semiclassical requirements of large action ${\cal R},
{\cal S} \gg \hbar$ and large spin angular momentum,~i.e.,
\be
  S \equiv \hbar s \gg \hbar.
\label{sinf}
\ee
Later the second condition will be dropped. The spin-orbit coupling is allowed
to be arbitrary now. Moreover, we show in Sec.\ \ref{lcac} that the WCL trace
formula \eq{trwclm} and the SCL trace formula \eq{trfSCL} can be reproduced
within the approach of \cite{plet}. 

In Sec.\ \ref{tfws} we presented the derivation of the spinless trace formula
starting from the quantum partition function. Here we apply the same scheme to
the spin-dependent $Z(T)$ of \eq{parti} with the c-number Hamiltonian \eq{chs}
responsible for the classical dynamics. At the end we should obtain the trace
formula in its standard form \eq{trform}. However, all its elements (periodic
orbits, their actions, periods, and stabilities) are to be found from the
dynamics in the extended phase space ${\bf q}, {\bf p}, z^*, z$. In what
follows we show how these ingredients can be determined. 

The first step is to obtain the classical equations of motion from the
variational principle $\delta {\cal R} [{\bf q}, {\bf p}, z^*, z] = 0$, where,
in particular, $z^*$ and $z$ are varied independently. These generalized
Hamilton equations are
\be
  \dot{q}_i = \frac{\partial {\cal H}}{\partial p_i} , \quad
  \dot{p}_i = -\frac{\partial {\cal H}}{\partial q_i},
\label{dotqp}
\ee
\be
  \dot{z}^*  = i \frac{(1+|z|^2)^2}{2S} \frac{\partial {\cal H}}{\partial z},
  \quad \dot{z} =- i\frac{(1+|z|^2)^2}{2S} \frac{\partial {\cal H}}{\partial
  z^*}. 
\label{dotz}
\ee
We emphasize that these equations describe the coupled dynamics of spin and
orbital degrees of freedom and collectively determine the periodic orbits in
the extended phase space. 

The second pair of equations are not redundant, as may appear. Actually, they
prove the expected property that $z^*$ is indeed the complex conjugate of $z$
along the classical paths. After we have established this, we can rewrite Eqs.\
\eq{dotz} in terms of the real variables  $v = \mathrm{Im}\, z^* = -
\mathrm{Im}\, z$ and $u = \mathrm{Re}\, z^* = \mathrm{Re}\, z$ as
\be
  \dot{v} = \frac{(1\!+\!|z|^2)^2}{4S} \frac{\partial {\cal H}}{\partial u} ,
  \quad
  \dot{u} = -\frac{(1\!+\!|z|^2)^2}{4S} \frac{\partial {\cal H}}{\partial v}.
\label{dotvu}
\ee
These two equations are equivalent to  
\be
  \dot{{\bf n}}= {\bf C} \times {\bf n},
\label{dotn}
\ee
with ${\bf n}$ given by \eq{clsp}. (As was already mentioned, the WCL equation
\eq{ztm} is of the same form as \eq{dotz}, if ${\cal H}$ is calculated along
the unperturbed orbits.) It is worth noting that the equations of motion
\eq{dotqp}, \eq{dotvu} can be formulated in terms of the generalized Poisson
bracket
\be
  \dot{{\cal F}} = \left\{ {\cal F}, {\cal H}  \right\} =  \frac{\partial {\cal
  F}}{\partial {\bf q}}\frac{\partial {\cal H}}{\partial{\bf p}} -
  \frac{\partial {\cal F}}{\partial {\bf p}}\frac{\partial {\cal
  H}}{\partial{\bf q}} + \frac{(1+u^2 + v^2)^2}{4S}  \left(\frac{\partial {\cal
  F}}{\partial v}\frac{\partial {\cal H}}{\partial u} - \frac{\partial {\cal
  F}}{\partial u}\frac{\partial {\cal H}}{\partial v}  \right)
\label{poisbra}
\ee
for any function ${\cal F}$ of the extended-phase-space coordinates.

According to \eq{princ}, the action along a periodic orbit is 
\be
  {\cal S}_{po} (E) = \oint_0^{T(E)} \biggl[ \frac12
  ({\bf p}\!\cdot\!{\bf\dot q} - {\bf q}\!\cdot\!{\bf\dot p})
  + 2S\,\frac{u \dot v - v \dot u}{1\!+\!u^2+\!v^2} \biggr] dt.
\label{perorbact}
\ee
The functional of the second variations ${\cal R}_{po}^{(2)}[\bgrk{\eta},T]$ is
still given by \eq{Rpo2} with 
\be
  \bgrk{\eta} = (\bgrk{\lambda},\nu,\bgrk{\rho},\xi)
              = \frac 1 {\sqrt{\hbar}} \left(\delta{\bf q},\,
              \frac{2\sqrt{S}\, \delta v}{1+u^2+v^2},\,
              \delta{\bf p},\,
              \frac{2\sqrt{S}\, \delta u}{1+u^2+v^2}\right)\!,
\label{eta}
\ee
as well as the $2(d+1)$-dimensional unit symplectic matrix ${\cal J}$ and
the second variation of the classical Hamiltonian \eq{chs}
\beq
  {\cal H}^{(2)} &=&  \frac{1}{2} \sum_{i,j} \left[\lambda_i \lambda_j
  \frac{\partial^2{\cal H}}{\partial q_i \partial q_j} + 2 \lambda_i \rho_j
  \frac{\partial^2{\cal H}}{\partial q_i \partial p_j} + \rho_i \rho_j
  \frac{\partial^2{\cal H}}{\partial p_i \partial p_j}\right] \nonumber\\
   &~& +\,  \frac{1}{2}(\nu^2+\xi^2)\,
         \left(-u\,C_1-v\,C_2+C_3\right) \nonumber\\
   &~& +\,  \sqrt{S}\,\frac{(1+|z|^2)}{2}
         \left(\nu\frac{\partial{\bf n}}{\partial v}
              +\xi\frac{\partial{\bf n}}{\partial u}\right) \cdot \sum_{j}
       \left(\lambda_j \frac{\partial {\bf C}}{\partial q_j}
       +\rho_j\frac{\partial{\bf C}}{\partial p_j} \right)\!.
\eeq
The stability matrices and the Maslov indices are determined by the linearized
dynamics $\dot{\bgrk{\eta}}={\cal J}\,\partial{\cal H}^{(2)}\!/
\partial{\bgrk{\eta}}$, as usual \cite{sugi}. 

There is, however, one component of the trace formula which is absent for
spinless systems. It is the quantum phase correction that emerges in the
semiclassical limit of  the spin-coherent-state path integral \cite{ston}.
Originally derived by Solari \cite{sola}, Kochetov~\cite{koch}, and Vieira and
Sacramento \cite{visa} for the semiclassical spin propagator, it is given~by
\be
  \varphi^{SK} (T) = \frac 1 2 \int_0^T B(t) dt,
\label{phiKS}
\ee
where
\be
  B(t) = \frac 1 2 \left[ \frac \partial {\partial z^*} \frac
 {\left(1 + |z|^2\right)^2} {2S} \frac {\partial {\cal H}} {\partial z} +
 \mathrm{c.c.} \right]
\ee
is evaluated along the classical path. Being purely of a kinematic nature, this
phase plays the role of a normalization \cite{koch}. For the Hamiltonian
\eq{chs} we find
\be
  B(t) = -u C_1 -v C_2 + C_3 = 2 A(t) - [(f(t))^* z + f(t) z^*].
\label{Bt}
\ee
In the semiclassical limit of the partition function \eq{parti} the
Solari-Kochetov phase is calculated along the periodic orbits. Since
$\varphi^{SK} \sim \hbar^0$, this phase survives the Fourier-Laplace
transform from $T$ to $E$ without affecting the stationary-phase condition and
thus enters the trace formula:
\be
 \delta g_{sc}(E) = \sum_{po} \mathcal A_{po}(E) 
                    \cos \left(\frac{1}{\hbar} \mathcal S_{po}(E) + 
		    \varphi^{SK}_{po}(E) - \frac{\pi}{2} \sigma_{po} \right).
\label{trfspin}
\ee

Note that, as follows from \eq{Bt}, $(1/2)\, \eta (E)$ in the WCL result
\eq{modfs} corresponds to the semiclassical Solari-Kochetov phase.

\subsection{Applicability conditions and coupling regimes}
\label{lcac}

We now will show that the semiclassical approach of the previous section is
\emph{not} restricted to large spin, provided that the Hamiltonian is linear in
spin. Thus, we will go beyond the formal limit \eq{sinf} and present the
arguments justifying this step.  We also identify the possible regimes of
weak, strong, and intermediate coupling that are unified within the
semiclassical treatment of the extended phase space. 

The trace formula \eq{trfspin} results from a stationary-phase evaluation of
the path integral \eq{parti} for the partition function. In the discretized
version one finds the stationary-phase points in the space of $(2d + 2) N$
variables $\mathbf{q}(t_l)$, $\mathbf{p}(t_l)$, $z(t_l)$, $l=1, \ldots , N$
[cf.~\eq{measure}]. Each stationary point corresponds to a periodic orbit that
satisfies the semiclassical equations of motion \eq{dotqp}, \eq{dotz}. In this
section we evaluate the path integral in a different, but equivalent, way.
Namely, we first integrate over the spin variables $z(t_l)$ by stationary
phase, keeping the orbital part fixed. The result, which consists of an
exponential and a prefactor, depends on $\mathbf{q}(t_l)$, $\mathbf{p}(t_l)$.
Then we integrate it over the orbital variables, again by stationary phase.
At the end we should obtain the same sum over the periodic orbits \eq{trfspin},
but the two-step calculation allows for a better control of the approximations
made.  

The described procedure is very similar to what was done in Sec.\ \ref{WCL},
starting with the representation \eq{seppf} for $Z(T)$. The difference is that
now ${\cal M}_s [{\bf q}, {\bf p};T]$ of \eq{Ms1} is evaluated by stationary
phase. A similar calculation was done in Ref.\ \cite{zait}, where the trace
formula for a spin evolving in the external magnetic field $\mathbf C (\mathbf
q (t), \mathbf p (t))$ was derived within the approach of the previous section
(including the Solari-Kochetov phase). The result is valid even if $(\mathbf q
(t), \mathbf p (t))$ is a non-classical path, as in the present case. Although
the final formula in \cite{zait} was written for spin $s=1/2$, it can be
straightforwardly generalized to arbitrary spin. Transforming it from the
energy to the time domain, we obtain the stationary-phase result for ${\cal
M}_s [{\bf q}, {\bf p};T]$ which \emph{coincides} with the quantum-mechanically
exact expression \eq{Msexact} for any value of spin. 

It is clear now that the stationary-phase evaluation of
\be
  Z (T) = \int \frac {{\cal D}[{\bf q}]{\cal D} [{\bf p}]} {(2\pi \hbar)^d}\;
  \exp \left\{ i {\cal R}_0 [\mathbf{q,p};T]/ \hbar \right\} \frac{\sin \{ (s +
  1/2)\, \eta\, [\mathbf{q,p};T]\}}{\sin \{ \eta \,[\mathbf{q,p};T]/2\}}  
\label{ZTgen}
\ee
yields the trace formula \eq{trfspin}. The stationary-phase condition is
fulfilled by the classical periodic orbits $(\mathbf{q}(t), \mathbf{p}(t))$
which satisfy the equations of motion \eq{dotqp}. We will also need the
representation for the partition function, in which the orbital part of
the path integral \eq{parti} is evaluated by stationary phase, producing
\beq
  Z (T) &=& \int {\cal D} \mu_s [z^* , z] \sum_{po} \exp \,\{i{\cal R}_0
  [\mathbf q_{po}, \mathbf p_{po}, z^*, z;T]/\hbar + is {\cal R}_1 [\mathbf
  q_{po}, \mathbf p_{po}, z^*, z;T]\}  \nonumber \\  &~& \times \int \! \frac
  {{\cal D} [\bgrk{\eta}_{\mathrm{orb}}]} {(2\pi)^d}\, \exp{\{i{\cal
  R}_{po}^{(2)} [\bgrk{\eta}_{\mathrm{orb}}, z^*, z; T]\}} ,
\label{ZTgen2}
\eeq
where the periodic orbits in the $(\mathbf{q,p})$ space are determined by the
Hamiltonian \eq{chs} for a given non-classical path $z(t)$. The integral of
second variations of the orbital variables is defined in \eq{zscl0}-\eq{H2orb}
with the full Hamiltonian \eq{chs} (the vector of small variations
$\bgrk{\eta}_{\mathrm{orb}}$ ought not to be confused with the functional $\eta
[\mathbf{q,p};T]$.) Again, applying the stationary-phase approximation to
\eq{ZTgen2}, we should recover the trace formula \eq{trfspin}. 

With the help of \eq{ZTgen} and \eq{ZTgen2} we can consider the effect of
spin-orbit interaction on the orbital degrees of freedom. Assuming that for the
orbital Hamilton's principal function the semiclassical condition $\mathcal
R_0/ \hbar \gg 1$ is always fulfilled and $\eta, {\cal R}_1 \sim |{\bf C}|\,
T_{\mathrm{orb}}$, we distinguish the following regimes:

(i) $\mathcal R_0/ \hbar \gg s \eta$ (WCL). The orbital classical dynamics is
determined by ${\cal H}_0$, i.e., the spin-dependent factor of \eq{ZTgen} does
not influence the stationary-phase condition. Thus, the results of Sec.\
\ref{WCL} are recovered. If, in addition, $|{\bf C}|\, T_{\mathrm{orb}}  \gg 1$
along the periodic orbit, $\eta$ is given by \eq{etaAdLim} and the fast spin
precession can be eliminated from the action. The same conclusions follow from
\eq{ZTgen2}, where the periodic orbits will not depend on the spin path $z(t)$,
and we can pull the periodic orbit sum and all the factors, except for $\exp
[is \mathcal R_1]$, out of the path integral.  

(ii) $\mathcal R_0/ \hbar \sim \eta$ (SCL). In this case both $\mathcal R_0$
and $\mathcal R_1$ determine the stationary-phase condition in \eq{ZTgen2}. The
fluctuation integral does not explicitly contain $\hbar$ and is slowly varying. 
Since the adiabatic condition $|{\bf C}|\, T_{\mathrm{orb}} \sim \mathcal
R_0/ \hbar  \gg 1$ is satisfied, the orbital motion is not sensitive to
small variations of the spin path $z(t)$. This means that the second-variation
integral after the stationary-phase integration over ${\cal D} \mu_s [z^* , z]$
will be determined only by the spin-dependent part of $\mathcal R_1$, i.e.,
\beq
  {\cal Z}_{sc}(T) &=& \sum_{po} \exp\{i{\cal R}_{0}/\hbar + i (s + 1/2)
  \eta\} \int \frac {{\cal D} [\bgrk{\eta}_{\mathrm{orb}}]} {(2\pi)^d}\,
  \exp{\{i{\cal R}^{(2)} [\bgrk{\eta}_{\mathrm{orb}}, z^*, z; T]\}} \nonumber
  \\
  &~& \times  \int \frac {{\cal D} [\bgrk{\eta}_{\mathrm{spin}}]} {2\pi}\,
  \exp{\{i{\cal R}_1^{(2)} [\mathbf q, \mathbf p, \bgrk{\eta}_{\mathrm{spin}};
  T]\}},
\label{ZSCL}
\eeq
where all the quantities are taken along the periodic orbits, and the integrals
of second variations are calculated separately for the orbital and the spin
variations. The Solari-Kochetov phase $\eta/2$ was added to the exponential.
The Fourier-Laplace transform \eq{gofE} of ${\cal Z}_{sc}(T)$, done by
stationary phase, yields the trace formula. The stationary-phase condition $E =
- \pd \mathcal R / \pd T$ seems to be spoiled by the Solari-Kochetov phase.
However, as shown in Appendix \ref{adext}, the periodic orbits form continuous
families on the Bloch sphere in the adiabatic limit. Therefore, $\eta$, being
the angle of rotation of the Bloch sphere (cf.\ Sec.\ \ref{WCL} and Ref.\
\cite{zait}), is a multiple of $2\pi$. Thus, $\eta$ remains constant as $T$ is
varied, i.e., $E = - \pd \mathcal R / \pd T = - \pd \mathcal R_0 / \pd T$ (this
does not mean, of course, that the function $E(T)$ is independent of spin,
since $\mathcal R_0$ is evaluated along the periodic orbit in the extended
phase space). After the trace formula is resummed to eliminate the spin part of
action and stability denominator (cf.\ Appendix \ref{adext}), the result
\eq{trfSCL} will be reproduced. Clearly, the orbital integral of second
variations in \eq{ZSCL} will be transformed to the stability determinants for
the periodic orbits of ${\cal H}_{\mathrm{eff}}^{(m_s)}$ \eq{Heff}. 

(iii) $\mathcal R_0/ \hbar \sim \eta$, but in some regions of phase space the
magnetic field $|{\bf C(q,p)}|$ becomes small, so that the adiabatic condition
fails (intermediate coupling). This is the regime when the semiclassical theory
in the extended phase space is especially useful, since both the WCL and SCL
formulations are not applicable. In particular, the general trace formula
\eq{trfspin} should remain valid if the orbit contains mode-conversion points
where ${\bf C(q,p)}=0$. In the weak-coupling regions the influence of spin
degrees of freedom on the stationary-phase point is negligible, of course. 

(iv) $S \sim \mathcal R_0 \gg \hbar$ (large-spin limit). This is the standard
case when the spin-phase-space semiclassics generally works, even if the
Hamiltonian is nonlinear in spin. The orbital motion is affected by spin, and
the trace formula \eq{trfspin} is appropriate here. 

We conclude that  \emph{for a Hamiltonian linear in spin, the general
semiclassical requirement of large spin $S \gg \hbar$, under which the trace
formula \eq{trfspin} was derived, now becomes unnecessary.} In fact, in many
physical problems the spin is small ($S \sim \hbar$) and the Hamiltonian is
linear in spin.  The trace formula \eq{trfspin} can be used in \emph{all four
regimes.} It can be simplified in the limits of weak and strong coupling, but,
since these are asymptotic limits, their boundaries may not be strictly defined
in specific numerical examples. The trace formula in its general form ensures
that all the effects of spin-orbit coupling on the density of states are taken
into account in the semiclassical approximation.\footnote{At this stage we have
to restrict ourselves to the case of ${\bf C(q,p)}$ depending on either $q_j$
or $p_j$, but not both, for each $j = 1, \ldots, d$, due to the problem of the
``no-name'' term (see Sec.\ \ref{AL}).}

It can be inferred from the discussion in Appendix \ref{adext} that the number
of periodic orbits in the extended phase space increases with the increase of
spin-orbit interaction strength. As a consequence, the number of bifurcations
may rise, making our theory, in a sense, technically more challenging. However,
we believe that the method should work reasonably well when the number of
relevant orbits is not too high, e.g., in the intermediate coupling regime (see
the example in Sec.~\ref{RashbaOscill}). An extension of our theory that
includes the treatment of bifurcations by uniform approximations is desirable.

As was already mentioned at the end of Sec.\ \ref{scsqpf}, there is a freedom
in choosing the c-number Hamiltonian for the path-integral representation of
the partition function~\eq{parti}. It is important, therefore, to understand
how the semiclassical theory of the previous section depends on a particular
choice of this Hamiltonian. In our path-integral construction we assigned to
the spin operator $\hbar \hat \mathbf s$ the classical symbol $\hbar s \mathbf
n$. The difference between this symbol and other possible symbols is $\mathcal
O (\hbar)$ (not $\mathcal O (\hbar s)$!). Thus, different symbols of the spin
Hamiltonian $\widehat{H}_s$ disagree by $\mathcal O (\hbar |\mathbf C|)$, which
is of the order of ${\cal H}_s$ itself, unless the spin is large. This means
that outside of the large-spin limit the results of Sec.\ \ref{seps} are valid,
generally speaking, only with the specific choice of the c-number Hamiltonian
\eq{chs}. Note that even in the large-spin limit, a change of the symbol will
show up in the phase of the trace formula, since the action will be modified by
$\mathcal O (\hbar |\mathbf C| T)$.

\section{Applications of semiclassical methods}
\label{asm}

We shall now apply the semiclassical methods of Sec.\ \ref{ssoc} to two
specific systems. In Sec.\ \ref{RJC} we study the free two-dimensional electron
gas with the Rashba Hamiltonian and the Jaynes-Cummings model, that allow
analytical treatment on both quantum-mechanical and semiclassical levels. Then
in Sec.\ \ref{RashbaOscill} we consider a numerical example of a quantum dot
with  harmonic confinement and  Rashba interaction. This system is a good test
case for our new semiclassical approach in the extended phase space, since the
SCL method suffers from the mode-conversion problem, while the WCL trace
formula completely neglects the spin-orbit interaction.

\subsection{Rashba and Jaynes-Cummings models}
\label{RJC}

The free two-dimensional electron gas with a Rashba spin-orbit interaction
\cite{rash} in a homogeneous magnetic field ${\bf B} = B_0 {\bf e}_3$ is
characterized by the Hamiltonian
\be
  \widehat{H} = \frac{\hat{\pi}_1^2}{2m^*} +\frac{\hat{\pi}_2^2}{2m^*}
  +\frac{2 \alpha_R}{\hbar} (\hat{\pi}_1 \hat{s}_2 - \hat{\pi}_2 \hat{s}_1
  ) + g^*  \mu_B B_0 \hat{s}_3.
\label{free2deg}
\ee 
Using the symmetric gauge for the vector potential, the components of
noncanonical momentum are given by $\hat{\pi}_1 = \hat{p}_1 - (e B_0 /2c)\,
\hat{q}_2$ and $\hat{\pi}_2 = \hat{p}_2 + (eB_0 / 2c)\, \hat{q}_1$. $\alpha_R$
is the Rashba constant \cite{rash,darn}, $m^*$ is the effective mass of an
electron, $e$ is the absolute value of its charge, $g^*$ is the effective
gyromagnetic ratio, $\mu_B = e \hbar / 2 m c$ is the Bohr magneton.

Due to the commutation relation $[ \hat{\pi}_2 , \hat{\pi}_1 ] = i \hbar e B_0
/c$ we can introduce canonically conjugated operators $\hat{Q} = -\hat{\pi}_2 
\sqrt{c / e B_0}$ and $\hat{P} = -\hat{\pi}_1 \sqrt{c /e B_0}$ satisfying  $[
\hat{Q} , \hat{P} ] = i \hbar$.  Then, the Hamiltonian \eq{free2deg} can be
written as
\be
  \widehat{H} = \frac{\omega_c}{2} (\hat{Q}^2+ \hat{P}^2 ) +2 \hbar  \kappa
  \omega_c (\hat{Q} \hat{s}_1 -  \hat{P}  \hat{s}_2) 
   + \hbar  \gamma \omega_c \hat{s}_3 ,
\label{free2deg3}
\ee
where we have introduced the new parameters\footnote{This definition of
$\kappa$ is applicable only within Sec.\ \ref{RJC}.} $\kappa = \alpha_R\,
\sqrt{eB_0} / \hbar^2 \sqrt{c}$, $\gamma = g^* m^* /2 m$, and the cyclotron
frequency $\omega_c= e B_0 /m^* c$. 

The Jaynes-Cummings model \cite{jacu}, that describes the simplest possible
interaction between a bosonic mode and a two-level system, has a similar
Hamiltonian, although all its ingredients have a different physical meaning. It
can be recovered after subtracting  $\hbar \omega_c /2$ from the Hamiltonian
\eq{free2deg3} and taking $s=1/2$.

For $s=1/2$ the quantum-mechanical energy spectrum of \eq{free2deg3} is known
analytically~\cite{rash}:
\be
  E_0 = \hbar\omega_c  (1 - \gamma) /2\,, \quad 
  E^\pm_k = \hbar\omega_c\left(k\pm\sqrt{(1 - \gamma )^2/4 
          +2k \hbar \kappa^2}\,\right), \quad k=1,2,3,\dots 
\ee
Using Poisson summation, it can be identically transformed to an exact
quan\-tum-mecha\-ni\-cal trace formula \cite{cham}. The smooth part is
$\widetilde g(E) = 2/ \hbar \omega_c$ and the oscillating part becomes
\beq
  \delta g(E) &= &\frac 2 {\hbar \omega_c} \sum_\pm \left( 1 \pm \frac {\hbar
  \kappa^2} {\sqrt {(1 - \gamma )^2/4 + 2E \kappa^2/ \omega_c + \hbar^2
  \kappa^4}} \right) \nonumber \\
  &\times& \sum_{k=1}^\infty \cos \left[ 2\pi k \left( \frac E {\hbar \omega_c}
  + \hbar \kappa^2 \pm \sqrt {(1 - \gamma )^2/4 + 2E \kappa^2/ \omega_c +
  \hbar^2 \kappa^4} \right) \right] . 
\eeq
One can also find analytically the oscillating part of the level density in the
WCL and SCL \cite{cham}.  

To illustrate the semiclassical methods discussed in this paper, we will
generalize $\delta g_{sc}^{\mathrm{WCL}}$ to arbitrary $s$, first, using the
WCL approach of Sec.\ \ref{WCL}, and, second, working in the extended phase
space (Sec.\ \ref{seps}) in the WCL.

\subsubsection{WCL method}

According to \eq{chs} the phase-space symbol of the Hamiltonian \eq{free2deg3}
is
\be
  {\cal H} = \frac{\omega_c}{2} (Q^2 + P^2) +2 \hbar s \kappa \omega_c ( 
  Q n_1 -  P n_2) + \hbar s \gamma \omega_c n_3 .
\label{free2deg4}
\ee
It is suitable to introduce the complex variables $\alpha = (Q + i P)/
\sqrt{2}$ and $n = n_1 + i n_2$ and to rewrite the Hamiltonian \eq{free2deg4}
in the form 
\be
  {\cal H} = \omega_c \alpha^* \alpha  + \sqrt{2} \hbar s \kappa
  \omega_c (\alpha n + \alpha^* n^* ) + \hbar s \gamma \omega_c n_3 .
\label{2deg}
\ee

In the WCL we can identify two small dimensionless parameters: $\hbar \omega_c
/E$ (validity of semiclassics) and $\hbar \kappa^2$ (weak coupling). We  impose
that $ \hbar \omega_c /E \sim \hbar \kappa^2 \ll 1$, meaning that in the double
limit $E \to \infty$, $\kappa \to 0$ the combination $E \kappa^2 / \omega_c$ is
kept constant. It corresponds to the formal limit $\hbar \to 0$.

We need to solve Eq.\ \eq{ztm}  with $A^{(0)} =  \gamma \omega_c /2$ and
$f^{(0)} =   \kappa \sqrt{2 E \omega_c}\, \exp (- i \omega_c t)$. Its periodic
solutions are
\be
  z = - |z| \exp (-i \omega_c t),
\ee
where $|z|$ is constant and is given by the quadratic equation
\be
  |z|^2 - \frac{1 - \gamma}{\kappa \sqrt{2 E \omega_c}} |z| -1 =0.
\ee
Of the two roots
\be
  |z|_{\pm} = \frac{\omega_c (1- \gamma)}{2 \kappa \sqrt{2 E \omega_c}} \pm
  \sqrt{\frac{\omega_c (1 - \gamma)^2}{8 E \kappa^2 } + 1} 
\ee
we will choose $|z|_{+}$ ($|z|_{-}$ yields the same result) and calculate the
integrand of \eq{stangm}
\be
  2 A^{(0)} -  (f^{(0) *} z + f^{(0)} z^*) = \omega_c (1 + 2 \Delta ),
\label{et} 
\ee
where
\be
  \Delta =  \sqrt{ \frac{(1 - \gamma )^2}{4}  + \frac{2 E \kappa^2}{\omega_c
  }}.
\label{stang}
\ee

The unperturbed trace formula for the system without spin degrees of freedom,
corresponding to ${\widehat H}_0=\hat{\bgrk{\pi}}^2\!/2m^*$, is that of a
one-dimensional harmonic oscillator with the cyclotron frequency $\omega_c$ and
reads \cite{book} 
\be
  \delta g^{(0)}(E)= \frac{2}{\hbar \omega_c}\sum_{r=1}^{\infty} (-1)^r \cos
  \left( r \frac{2\pi E}{\hbar\omega_c}\,\right).
\label{trace1ho}
\ee
Integrating \eq{et} over the $r$ repetitions of the primitive period of the
unperturbed system $T_{po}^{(0)} = 2 \pi / \omega_c$, we find
\be
  \eta_r (E) = 2 \pi r (1 + 2 \Delta ) .
\label{ert}
\ee
Then, according to \eq{trwclm}, the WCL trace formula is
\be
  \delta g_{sc}^{\mathrm{WCL}}(E)  = \frac{2}{\hbar \omega_c}
  \sum_{r=1}^{\infty} (-1)^{r (2s+1)} \frac{\sin [(2s +1) 2 \pi r \Delta
  ]}{\sin [2 \pi r \Delta ] } \cos \left( \frac{2 \pi E}{\hbar \omega_c } r
  \right) . 
\label{mmf}
\ee
For spin $s=1/2$ it yields 
\beq
  \delta g_{sc}^{\mathrm{WCL}}(E) &=& \frac{2}{\hbar \omega_c}
  \sum_{r=1}^{\infty} 2 \cos \left[ 2\pi r \Delta \right] \cos \left( \frac{2
  \pi E }{\hbar \omega_c } r \right) \nonumber \\
  &=& \frac{2}{\hbar\omega_c}\!\sum_\pm
  \sum_{r=1}^\infty \cos\!\left[ 2\pi r \left( \frac{E}{\hbar\omega_c}
  \pm\Delta \right)\!\right],
\label{tracewcl}
\eeq
in agreement with the result of \cite{cham}.

\subsubsection{WCL in the extended phase space}

Now we will carry out the procedure outlined in Sec.\ \ref{seps} for the Rashba
Hamiltonian. The implementation of the WCL in the extended phase space for a
general Hamiltonian of the form \eq{qh} can be found in Ref.\ \cite{zait}. 

The classical dynamics of the system is described by the equations of
motion \eq{dotqp}, \eq{dotn}, which now have the form
\beq
  &&\dot{\alpha} = -i\omega_c (\alpha +\sqrt{2} \hbar s \kappa n^*), 
  \nonumber \\
  &&\dot{n} = i\omega_c (\gamma n - 2\sqrt{2} \kappa n_3 \alpha^*),
  \label{eomr} \\
  &&\dot{n}_3 = -i \sqrt{2} \kappa \omega_c (n \alpha - n^*
  \alpha^*).\nonumber 
\eeq
Along with the energy conservation ${\cal H} = E$ there is another conserved
quantity $| \alpha |^2 +  \hbar s n_3 = \mathrm{const}$. It means that the
system \eq{free2deg4} is classically integrable. Its general analytic solution
has been given in \cite{grab}. Here, however, we are interested in the {\em
periodic} solutions, which infer the knowledge of certain initial conditions.
Let us make an ansatz for the particular isolated periodic orbits with $n_3 =
\mathrm{const}$:
\be
  \alpha = | \alpha | e^{-i \Omega t},  \quad n = -\xi |n| e^{i \Omega
  t}\quad  \quad (\xi = \pm 1 ) 
\label{ans} 
\ee
with constant parameters $\Omega$, $|\alpha |$ and $|n|$. Plugging
\eq{ans} into \eq{eomr} we obtain the relations between them:
\beq
  \Omega &=& \omega_c (1- \sqrt{2} \xi \hbar s \kappa | n | / | \alpha | ),
\label{rp1} \nonumber \\
  \Omega &=& \omega_c (\gamma + 2\sqrt{2} \xi \kappa n_3 | \alpha | / | n | ),
\label{rp2} \\
  E &=& \omega_c ( | \alpha |^2  + \hbar s \gamma n_3 - 2 \sqrt{2} \xi
  \hbar s \kappa | \alpha | |n| ). \nonumber 
\label{rp3}
\eeq
For the $r$th repetition of the orbits \eq{ans} we can calculate the action
\beq
  {\cal S}_{po}^r (E) &=& \oint_0^{T(E)} dt \left( \frac {\alpha \dot{\alpha}^*
  - \alpha^* \dot{\alpha} } {2i} + \hbar s \frac{z \dot{z}^* - z^* \dot{z}}{i
  (1 + |z|^2 )} \right) = 2 \pi r (|\alpha|^2 + \hbar s n_3) + 2 \pi r \hbar s
  \nonumber \\
  &=& 2 \pi r \left( \frac{E}{\omega_c} + \hbar s (1 - \gamma) n_3 + 2
  \sqrt{2} \xi \hbar s \kappa | \alpha | |n| \right) + 2 \pi r \hbar s
\label{actrash}
\eeq
and the stability angle
\be
  \Lambda_{po}^r = r \frac{4 \pi \kappa \omega_c}{\Omega}
  \sqrt{\frac{\hbar^2 s^2 |n|^2}{2 | \alpha |^2} + \frac{2 | \alpha |^2}{|n|^2}
  - 2 \hbar s n_3 } + 2 \pi r.
\label{stabang}
\ee
 
It seems to be difficult to solve the algebraic equations  \eq{rp2} with
respect to $\Omega$, $|\alpha |$, and $|n|$. Moreover, from the numerical
search of the periodic orbits in this system we know that for rather large
$\kappa$ there might appear other periodic orbits with more complicated shapes.
Up to now we have not used the WCL conditions in our calculations. In
particular, the expressions for the action and stability angle are valid for
any values of parameters. We gain considerable simplifications in the WCL, that
corresponds to the formal expansion in a series of $\hbar$: $| \alpha | = |
\alpha |^{(0)} + \hbar | \alpha |^{(1)} + \ldots$, etc. Then we can find that,
for instance,
\beq
  &&| \alpha |^{(0)} = \sqrt{E / \omega_c} , \quad \Omega^{(0)} = \omega_c , \\
  &&n_3^{(0)} = \xi (1 - \gamma) / 2 \Delta , \quad
  |n|^{(0)} = \kappa \sqrt{2 E} / \Delta \sqrt{\omega_c } . 
\eeq
We conclude that, in the leading order, the orbital motion is unaffected by
spin and for the two periodic orbits we have $\mathbf n_{\,\xi = 1} (t) = -
\mathbf n_{\,\xi = -1} (t)$, i.e., the spins are opposite at any time.
Moreover, we know from the general considerations \cite{zait} that in the WCL
only two periodic orbits are possible for the Hamiltonian \eq{2deg}, and
they are given by \eq{ans}. For these orbits we can calculate the action ${\cal
S}_{\xi}^r = (2 \pi\, E / \omega_c) + \xi\, \hbar\, s\, \eta_r (E)$
[Eq.~\eq{ert}] and the stability angle $\Lambda_{\xi}^r = \Lambda^{(0)}_r =
\eta_r (E)$ using \eq{actrash} and \eq{stabang}, respectively. As was discussed
in Sec.\ \ref{seps}, the Solari-Kochetov extra phase $\varphi^{SK}_{\xi}$,
which in the WCL is equal to $(1/2)\, \xi\, \eta_r (E)$, should be added to the
trace formula. Thus, the total phase that enters the trace formula is (up to
the Maslov indices)
\be
  \Phi_{\xi}^r = \frac{{\cal S}_{\xi}^r}{\hbar} + \varphi^{SK}_{\xi} = \frac{2
  \pi r E}{\hbar \omega_c } + (2s + 1) 2 \pi r \xi\, (2 \Delta + 1) + O
  (\hbar).
\ee
Then we find the oscillating part of the level density
\be
  \delta g_{sc}^{\mathrm{WCL}} (E) = \frac{1}{\hbar \pi} \sum_{\xi
  =\pm 1}\sum_{r=1}^{\infty} \frac{ T^{(0)} }{2 |\sin (\Lambda^{(0)}_r /2)|}
  \cos (\Phi_{\xi}^r - \sigma_{\xi}^r \pi /2 ) ,
\label{ldrash}
\ee
where $T^{(0)}= 2 \pi /\omega_c$ is the unperturbed primitive period. Adding
the appropriate Maslov indices \cite{zait} $\sigma_{\xi}^r = 1 + 2\, [\xi \,
\eta_r (E)/ 2 \pi]$, where $[x]$ is the largest integer $\leq x$, and summing
up $\xi=\pm 1$, we can recover formula \eq{mmf}.

\subsection{Quantum dot with the Rashba interaction}
\label{RashbaOscill}

We consider a two-dimensional electron gas in a semiconductor heterostructure,
laterally confined to a quantum dot by a harmonic potential. We assume that its
Hamiltonian  
\be 
  \hat{H} = \frac{\hat{p}_1^2}{2 m^*} + \frac{\hat{p}_2^2}{2 m^*} + \frac{m^*\,
  \Omega_1^2\, \hat{q}_1^2}{2} + \frac{m^*\, \Omega_2^2\, \hat{q}_2^2}{2} + 2
  \kappa \hbar\, (\hat{s}_2 \hat{p}_1 - \hat{s}_1 \hat{p}_2) 
\label{qham} 
\ee
includes a spin-orbit interaction of Rashba type \cite{rash}, where\footnote
{This definition of $\kappa$ is applicable only within Sec.\
\ref{RashbaOscill}} $\kappa = \alpha_R/ \hbar^2$. The c-number Hamiltonian
\eq{chs} in this case is 
\be
  {\cal H} = \frac{p_1^2}{2 m^*} + \frac{p_2^2}{2 m^*} + \frac{m^* \Omega_1^2
  q_1^2}{2} + \frac{m^* \Omega_2^2 q_2^2}{2} + 2 S \kappa (n_2 p_1 - n_1 p_2).
\ee

We will treat this system within our semiclassical approach in the extended
phase space. As discussed in \cite{cham}, the WCL trace formula fails to
account for the spin-orbit interaction. Indeed, without spin-orbit coupling the
only periodic orbits of the system are the two self-retracting librations along
the principal axes (we assume $\Omega_1 / \Omega_2$ to be irrational).  The
effective magnetic field $\mathbf C (\mathbf p) = 2 \kappa \mathbf e_3 \times
\mathbf p$ changes its sign together with $\mathbf p$. Hence the spin
precession generated by a libration is self-retracting as well, making the
rotation angle for the Bloch sphere $\eta (E)$ vanish. Then the modulation
factor ${\cal M}_{po} (E) = 2s + 1$ for both orbits is trivial. Thus, the WCL
trace formula is that of the two-dimensional anisotropic harmonic oscillator
\cite{book} multiplied by the spin degeneracy factor~$(2s + 1)$: 
\beq
  \delta g_{sc}^{\mathrm{WCL}} (E) &=& \frac {2s + 1} {\hbar \Omega_1}
  \sum_{k=1}^{\infty} \frac {(-1)^k} {\sin \left( k \pi \frac {\Omega_1}
  {\Omega_2} \right)} \sin \left(k \frac {2\pi E} {\hbar \Omega_1} \right) 
  \nonumber \\
  &+& \frac {2s + 1} {\hbar \Omega_2} \sum_{k=1}^{\infty} \frac {(-1)^k} {\sin
  \left( k \pi \frac {\Omega_2} {\Omega_1} \right)} \sin \left(k \frac {2\pi E}
  {\hbar \Omega_2} \right).
\eeq

In the SCL the two pendulating orbits are still present and contain the
mode-conversion points \cite{cham}, where the interaction term vanishes. Thus
the SCL approach in its original form cannot be applied either. Although one
can fix it to some extent by an \emph{ad hoc} procedure \cite{frgu}, it would
be only natural for us to resort to our extended-phase-space semiclassics, that
does not suffer from the mode conversion problem and is not restricted to weak
or strong spin-orbit coupling.

The classical dynamics is described by the equations of motion \eq{dotqp} and
\eq{dotvu} that now become
\beq
  &~&\dot{q}_1 = \frac{p_1}{m^*} +\frac{4 S \kappa v}{1 + u^2 + v^2} =
  \frac{p_1}{m^*} + 2 S \kappa n_2, \\
  &~&\dot{q}_2 = \frac{p_2}{m^*} -\frac{4 S \kappa u}{1 + u^2 + v^2} =
  \frac{p_2}{m^*} - 2 S \kappa n_1, \\
  &~&\dot{v} = - \kappa (1+ u^2 + v^2) p_2 + 2 \kappa u ( u p_2 -v p_1 ),
  \label{eqv} \\
  &~&\dot{p}_1 = - m^* \Omega_1^2 q_1 , \\
  &~&\dot{p}_2 = - m^* \Omega_2^2 q_2 , \\ 
  &~&\dot{u} = - \kappa (1+ u^2 + v^2) p_1 - 2 \kappa v ( u p_2 -v p_1 ).
  \label{equ}
\eeq
Equations \eq{eqv} and \eq{equ} are equivalent to \eq{dotn}, or,
\be
  \dot{n}_1 = 2 \kappa p_1 n_3 , \quad
  \dot{n}_2 = 2 \kappa p_2 n_3 , \quad
  \dot{n}_3 = - 2 \kappa (p_1 n_1 + p_2 n_2 ) .
\ee

For a numerical study, it is convenient to describe the system in dimensionless
quantities. Let us choose a characteristic frequency $\Omega_0$ and define the
dimensionless $\omega_1 = \Omega_1 / \Omega_0$ and $\omega_2 = \Omega_2 /
\Omega_0$ (note that there is a degree of arbitrariness in the choice of
$\Omega_0$). Then we can construct the following scaled variables:
\beq
  &~& \tilde{{\bf q}} = {\bf q} \sqrt{\frac{m^* \Omega_0}{2 S}} , \quad \tilde
  {\bf p}  = {\bf p} \frac 1{\sqrt{2 S m^* \Omega_0}} , \quad \tilde{\kappa} =
  \kappa \sqrt{\frac{2S   m^*}{\Omega_0}} , \\  
  &~& (\tilde{E}, \tilde {\cal H}) = \frac 1 {2 S \Omega_0} (E, {\cal H}),
  \quad \tilde{t} =  \Omega_0 t, \quad (\tilde {\cal S}, \tilde {\cal R}) =
  \frac 1 {2 S} ({\cal S}, {\cal R}).
\eeq
The Hamiltonian in the scaled coordinates has the form
\be
  \tilde{{\cal H}} = \frac{\tilde{p}_1^2}{2} + \frac{\tilde{p}_2^2}{2} +
  \frac{\omega_1^2 \tilde{q}_1^2}{2} + \frac{\omega_2^2 \tilde{q}_2^2}{2} + 2
  \tilde{\kappa} \frac{ v \tilde{p}_1 - u \tilde{p}_2 }{1 + u^2 + v^2} .
\label{sclH}
\ee
With the scaled action \eq{perorbact}
\be
  \tilde {\cal S} = \oint \left[ \frac12 (\tilde{{\bf p}} \cdot d \tilde{{\bf
  q}} - \tilde{{\bf q}} \cdot d \tilde{{\bf p}}) + \frac{u dv - v du}{1 + u^2 +
  v^2} \right]
\ee
the applicability condition of the semiclassical approach now reads
$\tilde{{\cal S}} \gg 1/ 2s$.

The equations of motion for the tilded variables  are (time derivatives with
respect to~$\tilde{t}$)
\beq
  \dot{\tilde{q}}_1 &=& \tilde{p}_1 + \tilde{\kappa} n_2, \quad
  \dot{\tilde{q}}_2  = \tilde{p}_2 - \tilde{\kappa} n_1, 
\label{em1}\\
  \dot{\tilde{p}}_1 &=& - \omega_1^2 \tilde{q}_1 , \quad
  \dot{\tilde{p}}_2  =  - \omega_2^2 \tilde{q}_2 , 
\\
  \dot{n}_1 &=& 2 \tilde{\kappa}  \tilde{p}_1 n_3 , \quad
  \dot{n}_2  =  2 \tilde{\kappa}  \tilde{p}_2 n_3 , \quad
  \dot{n}_3  =- 2 \tilde{\kappa} (\tilde{p}_1 n_1 + \tilde{p}_2 n_2 ) .
\label{em3}
\eeq
Note that the system possesses certain discrete symmetries:
\beq
  R_{12} &:& \tilde{q}_1 \to -\tilde{q}_1 , \quad  \tilde{q}_2 \to
  -\tilde{q}_2, \quad  n_3  \to  - n_3, \quad \tilde{t} \to -\tilde{t} \\
  R_{1} &:& \tilde{q}_1 \to -\tilde{q}_1 , \quad  \tilde{p}_2 \to -\tilde{p}_2
  , \quad n_1  \to  - n_1 , \quad \tilde{t} \to -\tilde{t} \\ 
  R_{2} &:& \tilde{q}_2 \to -\tilde{q}_2 , \quad  \tilde{p}_1 \to -\tilde{p}_1
  , \quad n_2  \to  - n_2 , \quad \tilde{t} \to -\tilde{t} 
\eeq

There are trivial solutions to the equations of motion---the pendulating orbits
with ``frozen spin'':
\begin{itemize}
\item  The pair of orbits $A_x^\pm$ pendulating along the $\tilde q_1$ axis
with spin $n_2 = \pm 1$. The phase-space coordinates along these orbits are 
\beq
  &~&\tilde{q}_1 (\tilde{t}) = \pm \left(\sqrt{2 \tilde{E} + \tilde{\kappa}^2}
  / \omega_1 \right) \sin (\omega_1 \tilde{t}), \quad \tilde{p}_1 (\tilde{t}) =
  \mp \tilde{\kappa} \pm  \sqrt{2 \tilde{E} + \tilde{\kappa}^2}  \cos (\omega_1
  \tilde{t}), \nonumber \\
  &~&v(\tilde{t}) = \pm 1, \quad \tilde{q}_2 (\tilde{t}) = \tilde{p}_2
  (\tilde{t}) = u (\tilde{t}) = 0. 
\eeq
The period of the orbits is $\tilde{T}_1 = 2 \pi / \omega_1$. The orbits are
invariant under $R_{12}$ and $R_{1}$ ($R_{2}$ produces the symmetry partner,
i.e., maps $+$ onto $-$).
\item Similarly, there are a pair of orbits $A_y^\pm$ pendulating in the
$\tilde q_2$ direction with spin  $n_1 = \pm 1$. They are given by 
\beq
  &~&\tilde{q}_2 (\tilde{t}) = \pm \left(\sqrt{2 \tilde{E} + \tilde{\kappa}^2}
  / \omega_2 \right) \sin (\omega_2 \tilde{t}), \quad \tilde{p}_2 (t) = \pm
  \tilde{\kappa}  \pm \sqrt{2 \tilde{E} + \tilde{\kappa}^2} \cos (\omega_2
  \tilde{t}), \nonumber \\
  &~&u(\tilde{t}) = \pm 1, \quad \tilde{q}_1 (\tilde{t}) = \tilde{p}_1
  (\tilde{t}) = v (\tilde{t}) = 0. 
\eeq
The period is $\tilde{T}_2 = 2 \pi /\omega_2$. The orbits are  invariant under
$R_{12}$ and $R_{2}$ ($R_{1}$ produces the symmetry partner). 
\end{itemize}

\begin{figure}[t]
  \vspace*{.5cm}
  \begin{center}
    {\hspace*{0cm}
    \psfig{figure=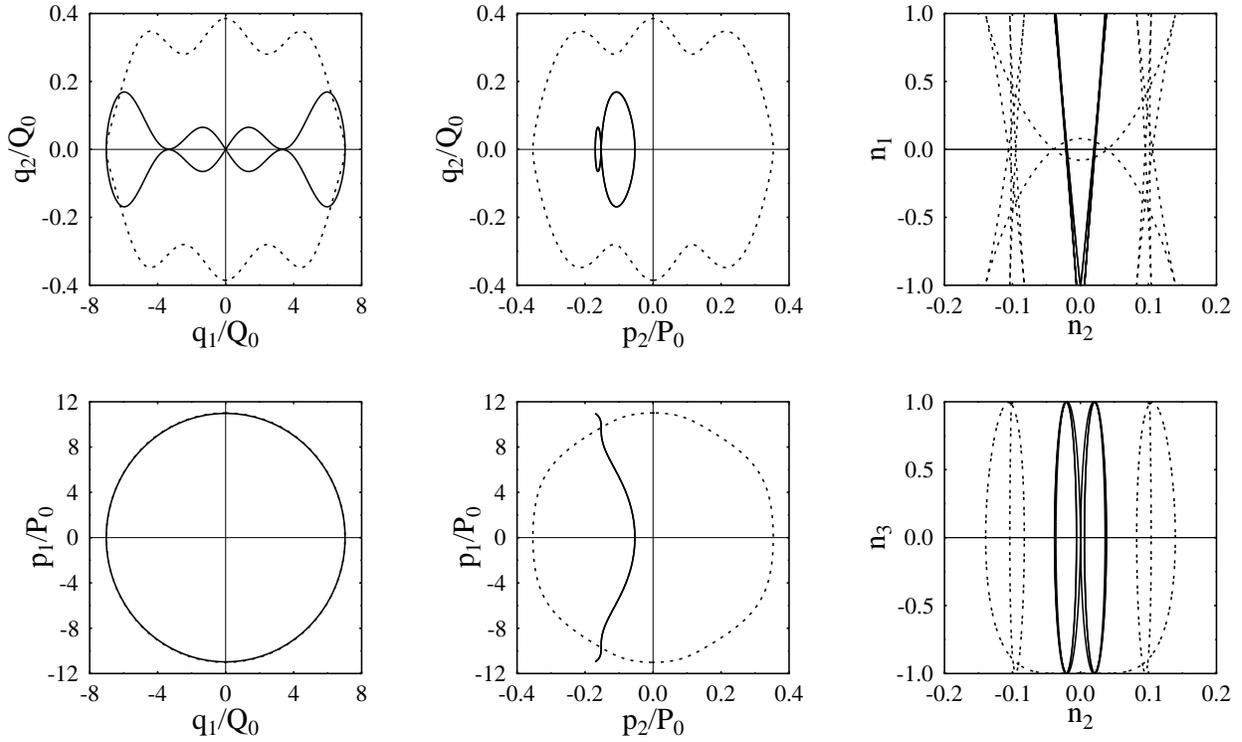,width= \textwidth}}
  \end{center}
  {\vspace*{0cm}}
  \caption{{\small Periodic orbits $D_{x1}^+$ (solid line) and $D_{x2}^+$
  (dotted line) for $\tilde \kappa = 0.67$, $\tilde E = 60$, represented in
  different cross-sections of the phase space. Here $Q_0 = \sqrt{2 S / m^*
  \Omega_0}$, $P_0 = \sqrt{2 S m^* \Omega_0}$. Note the different scales along
  the axes. In the lower left plot the two orbits are almost indistinguishable.
  For the $\mathbf n (t)$ time dependence see \cite{plet}.}}
\label{Dx}
{\vspace{.5 cm}}
\end{figure}

\begin{figure}[t]
  \vspace*{.5cm}
  \begin{center}
    {\hspace*{0cm}
    \psfig{figure=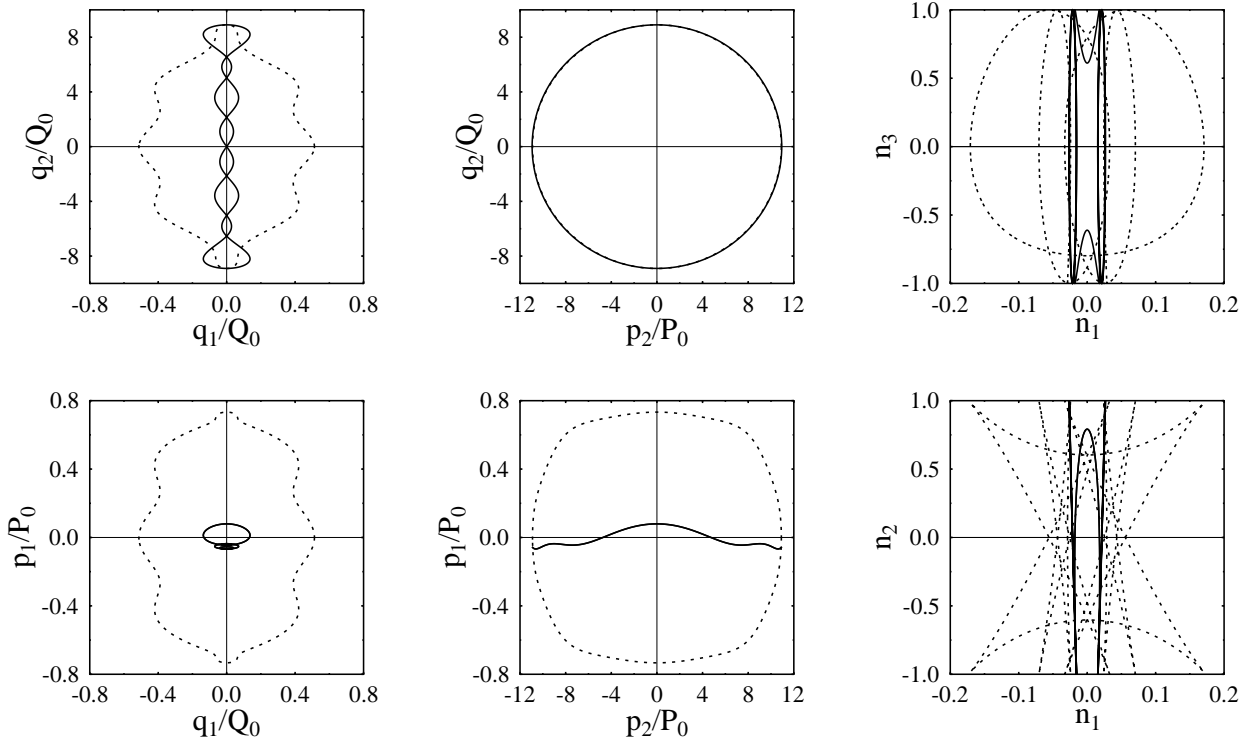,width= \textwidth}}
  \end{center}
  {\vspace*{0cm}}
  \caption{{\small Same as in Fig.\ \ref{Dx} for orbits $D_{y1}^+$ (solid line) and
  $D_{y2}^+$ (dotted line). In the
  upper middle plot the two orbits are almost indistinguishable.}} 
\label{Dy}
{\vspace{.5 cm}}
\end{figure}

\begin{figure}[t]
  \vspace*{.5cm}
  \begin{center}
    {\hspace*{0cm}
    \psfig{figure=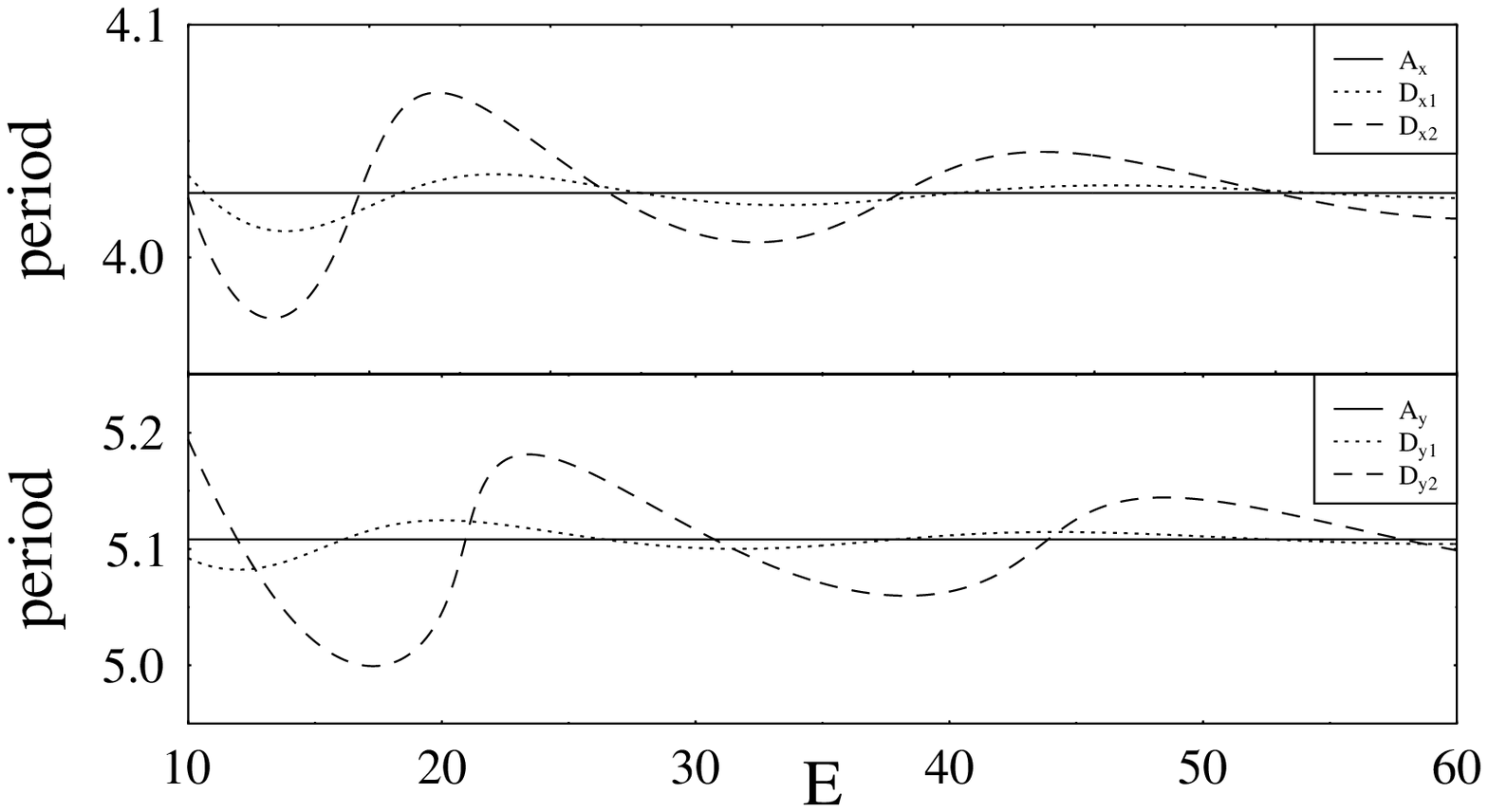,width= \textwidth}}
  \end{center}
  {\vspace*{0cm}}
  \caption{{\small The periods of the $A^\pm$ and $D^\pm$ orbits as functions
  of energy for $\tilde \kappa = 0.67$. The periods are given in units of
  $1/\Omega_0$, the energy unit is $2 S \Omega_0$. For the orbits $A_x^\pm$ and
  $A_y^\pm$ the periods are independent of energy and equal to $\tilde{T}_1
  \approx 4.02768$ and $\tilde{T}_2 \approx 5.10828$, respectively.}} 
\label{per}
{\vspace{.5 cm}}
\end{figure}

Other types of periodic orbits can be found from the numerical solution of the
equations of motion \eq{em1}-\eq{em3}. In our numerical example we use
$\omega_1 = 1.56$ and $\omega_2 = 1.23$. For a large range of parameters with
$0 < \tilde \kappa \lesssim 0.75$ and $\tilde E \gtrsim 8$ we find the
following non-trivial periodic orbits:
\begin{itemize}
\item  Two pairs of orbits $D_{x1}^\pm$ and $D_{x2}^\pm$ oscillating around
$A_x^\pm$ in the configuration space, with $n_2 \sim 0$ (Fig.\ \ref{Dx}). The
spin is rotating about $n_2$ axis. The superscripts ($\pm$) denote the sense of
rotation in the subspace $(\tilde q_1, \tilde q_2)$.
\item  Two pairs of orbits $D_{y1}^\pm$ and $D_{y2}^\pm$ oscillating around
$A_y^\pm$, with $n_1 \sim 0$ (Fig.\ \ref{Dy}). The spin is rotating about $n_1$
axis.
\end{itemize}
For stronger couplings $\tilde \kappa \gtrsim 0.75$ or smaller energies $\tilde
E \lesssim 8$, new orbits bifurcate from the $A$ and $D$ orbits. Near the
bifurcations the trace formula would have to be modified by uniform
approximations \cite{ssun}. The periods of the orbits are shown in Fig.\
\ref{per}.

The spin-orbit strength $\tilde \kappa$ depends on the band structure
\cite{darn}. For example, for an InGaAs-InAlAs quantum dot with $\sim 100$
confined electrons one would obtain a value of $\tilde \kappa \sim 0.25$. In
order to have the effect of spin on the orbital motion more pronounced, we
choose $\tilde \kappa = 0.67$ for our numerics. 

The stability determinant in the trace formula can be calculated according to
the general prescription (Sec.\ \ref{seps}). The second variation of the
Hamiltonian \eq{sclH} is
\beq 
  \tilde {\cal H}^{(2)} &=&  \frac {\tilde \rho_1^2} {2} + \frac {\tilde
  \rho_2^2} {2} + \frac {\omega_1^2 \tilde \lambda_1^2} {2} + \frac {\omega_2^2
  \tilde \lambda_2^2} {2}
\nonumber \\
  &+&   \tilde \kappa (u \tilde p_2 - v \tilde p_1 ) (\xi^2 + \nu^2) + \tilde
  \kappa  \frac {2 \sqrt 2\, u v} {1 + u^2 + v^2} (\tilde \rho_2 \nu - \tilde
  \rho_1 \xi ) 
\nonumber  \\
  &+& \tilde \kappa \sqrt 2 \left( \frac {2 u^2} {1 + u^2 + v^2} - 1 \right)
  \tilde \rho_2 \xi - \tilde \kappa \sqrt 2 \left( \frac {2 v^2} {1 + u^2 +
  v^2} -  1 \right)  \tilde \rho_1 \nu,
\eeq
where the variables $\bgrk \lambda$ and $\bgrk \rho$ are scaled like $\bf q$
and $\bf p$, respectively. Numerically solving the equations of motion for
variations
\beq
  \dot {\tilde \lambda}_1 &=& \tilde \rho_1 - \tilde \kappa \sqrt 2 \frac {2 u
  v} {1 + u^2 + v^2} \xi - \tilde \kappa \sqrt 2 \left( \frac {2 v^2} {1 + u^2
  + v^2} - 1 \right) \nu, 
\\
  \dot {\tilde \lambda}_2 &=& \tilde \rho_2 + \tilde \kappa \sqrt 2 \left(
  \frac {2 u^2} {1 + u^2 + v^2}- 1 \right) \xi +\tilde \kappa \sqrt 2 \frac
  {2uv} {1 + u^2 + v^2} \nu, 
\\
  \dot{\nu} &=& -\tilde \kappa \frac {2 \sqrt 2\, u v} {1 + u^2 + v^2} \tilde
  \rho_1 + \tilde \kappa \sqrt 2 \left( \frac {2 u^2} {1 + u^2 + v^2} - 1
  \right) \tilde \rho_2  + 2 \tilde \kappa (u \tilde p_2 - v \tilde p_1 ) \xi, 
\\
  \dot{\tilde \rho}_1 &=& - \omega_1^2 \tilde \lambda_1, 
\\
  \dot{\tilde \rho}_2 &=& - \omega_2^2 \tilde \lambda_2, 
\\
  \dot{\xi} &=& \tilde \kappa \sqrt 2 \left( \frac {2 v^2} {1 + u^2 + v^2} - 1
  \right) \tilde \rho_1 - \tilde \kappa \frac {2 \sqrt 2\, u v} {1 + u^2 + v^2}
  \tilde \rho_2  - 2 \tilde \kappa (u \tilde p_2 - v \tilde p_1 ) \nu  ,
\eeq
one determines the reduced monodromy matrix and then finds the stability
determinant $\det ( \widetilde M_{po} - I_4)$ of the periodic orbits. 

\begin{figure}[p]
  \vspace*{.5cm}
  \begin{center}
    {\hspace*{0cm}
    \psfig{figure=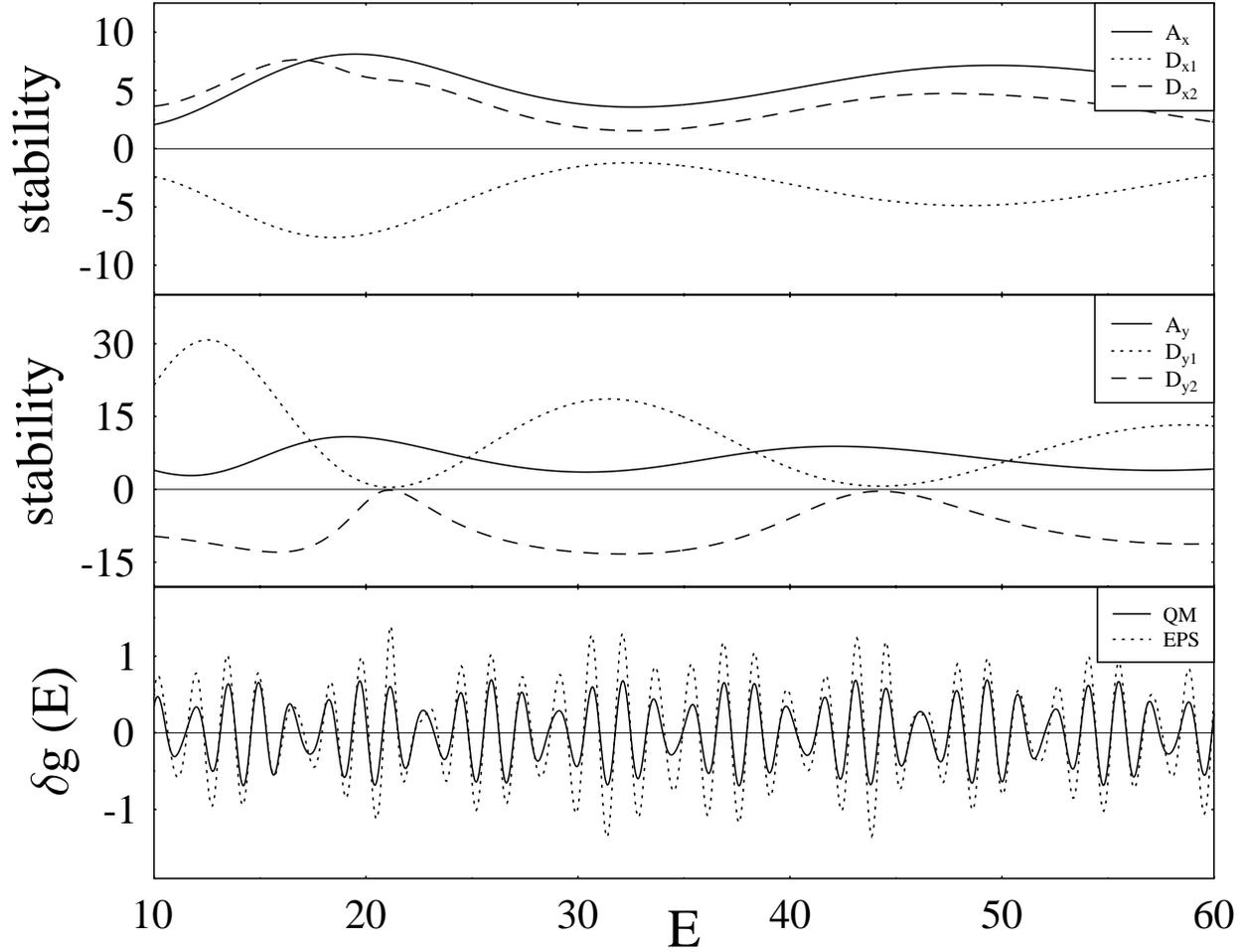,width= \textwidth}}
  \end{center}
  {\vspace*{0cm}}    
  \caption{{\small \emph{Upper and middle panels:} the stabilities $\det (
  \widetilde M_{po} - I_4)$ of the $A^\pm$ and $D^\pm$ orbits. The curves come
  close to the zero level, but do not touch or cross it (avoided bifurcations).
  \emph{Lower panel:} the oscillating part of density of states for $\tilde
  \kappa = 0.67$ and $S= \hbar / 2$. QM: the quantum-mechanical direct
  diagonalization. EPS: the semiclassical calculation in the extended phase
  space, including only the first repetitions of the 12 primitive periodic
  orbits $A$ and $D$ (Gaussian averaging parameter $\gamma = 0.6$). The energy
  is measured in units of $\hbar \Omega_0$. }} 
\label{stabdos}
{\vspace{.5 cm}}
\end{figure}

After calculating the Maslov indices within Sugita's approach \cite{sugi} and
the Solari-Kochetov phase by \eq{phiKS}, we can compute the oscillating part of
the density of states using the trace formula \eq{trfspin} (Fig.\
\ref{stabdos}). Note that while the classical dynamics in the scaled variables
is independent of the value of spin, the density of states will depend on $S$,
since the unscaled action $2S \tilde \mathcal S$ enters the phase of the trace
formula. We choose the physically meaningful $S= \hbar / 2$ in our example. To
ensure the convergence of the periodic orbit sum, the density of states was
convoluted with a normalized Gaussian, $\exp[ -(E/\gamma)^2] /\gamma \sqrt
\pi$, i.e., it was smoothed out with the energy window $\sim \gamma$. With the
averaging parameter $\gamma = 0.6$, the first repetitions of the 12 primitive
periodic orbits $A$ and $D$ were sufficient in the trace formula. The
semiclassical result for the density of states is compared with the
quantum-mechanical curve, obtained from a numerical diagonalization of the
Hamiltonian \eq{qham}. We observe a rather good agreement between the two,
especially for the oscillation frequencies. The difference in the amplitudes
can be explained by the vicinity of the bifurcations in the parameter space.
Indeed, the disparity becomes larger near the avoided bifurcations, where the
stabilities are extremely small (Fig.\ \ref{stabdos}). In principle, these
energy regions should be treated by a uniform approximation \cite{ssun}. The
matter is complicated, however, by the fact that the avoided bifurcations are
non-generic and of codimension larger than 1. The theory for such bifurcations
is developed only for two-dimensional systems. Our system is effectively
three-dimensional. In addition to the elliptic and (inverse) hyperbolic orbits,
it also has the loxodromic orbits. Thus, the extension of the standard theory
of bifurcations is difficult and still needs to be developed.

\begin{figure}[t]
  \vspace*{.5cm}
  \begin{center}
    {\hspace*{0cm}
    \psfig{figure=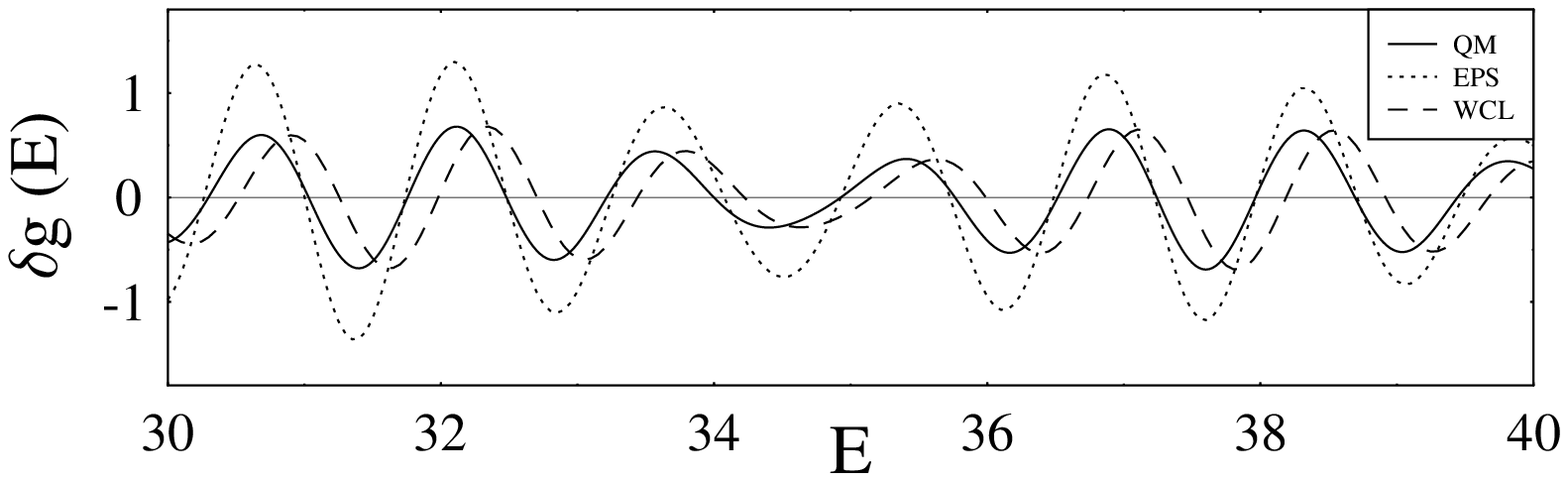,width= \textwidth}}
  \end{center}
  {\vspace*{0cm}}    
  \caption{{\small A detailed view of the lower panel of Fig.\ \ref{stabdos}
  with the WCL density of states added.}} 
\label{doszoom}
{\vspace{.5 cm}}
\end{figure}

A more detailed view of $\delta g (E)$ is shown in Fig.\ \ref{doszoom}. For
comparison, we added there the density of states calculated by the WCL trace
formula with spin $S= \hbar / 2$, which, in this case, is the trace formula of
the two-dimensional anisotropic harmonic oscillator without spin-orbit coupling
\cite{book} multiplied by the spin degeneracy factor 2 (see above). Clearly, it
is shifted by phase from the exact density of states. It is worth mentioning
that the dashed (WCL) curve would come very close to the exact
quantum-mechanical curve if the former is shifted by $-\tilde \kappa^2 / 2$ in
the scaled energy $\tilde E$. Note that the action of the trivial periodic
orbits is $\tilde \mathcal S_i = (\pi/ \omega_i) (2 \tilde E + \tilde
\kappa^2)$. Thus the shift can be related to the perturbative correction to the
action in the extended phase space, which is of the second order in $\tilde
\kappa$. To explain the shift, it would be interesting to develop the
second-order perturbation theory in the parameter $\tilde \kappa$ for the trace
formula in the extended phase space, similar to that proposed in \cite{brpt}.
(The first-order perturbation theory \cite{crpt} should be equivalent to the
WCL trace formula.) Apparently, the shift of the dashed curve cannot be
justified within the standard WCL approach of \cite{boke}. With the increase of
$\tilde \kappa$, the non-integrability of the system will show up, and the
shape of the WCL curve (describing the integrable system) will start to deviate
from the exact density of states. 

\section{Conclusion}

We have presented a detailed discussion of a new semiclassical method for
systems with spin-orbit coupling. The key idea of this approach is to introduce
an extended phase space of orbital and spin degrees of freedom where the
semiclassical dynamics takes place. The recipe for the construction of such a
phase space, as well as the equations of motion there, can be obtained from the
path-integral formulation of quantum mechanics using the spin coherent states,
with a subsequent stationary-phase evaluation. 

We have classified the possible regimes according to the strength of the spin-orbit
coupling and the value of the spin. When the Hamiltonian is linear in the spin, our
method works not only in the standard semiclassical domain of large spin, but
also for any finite spin. From the path-integral formulation we have directly
derived the trace formulae in the limiting cases of weak and (under certain
restriction) strong spin-orbit coupling for arbitrary spin. It was argued that
the semiclassical approach in the extended phase space recovers the limiting
behavior. 

The general method was illustrated with the specific examples relevant, in
particular, to the mesoscopic physics of heterostructures. Our analytical and
numerical results underscore the importance of the Solari-Kochetov phase
correction in the coherent-state path integrals and the proper evaluation of
the Maslov indices in the extended phase space. 

The future work in this direction might include the improvement of the
semiclassical evaluation of path integrals in the case of strong spin-orbit
coupling (restoration of the no-name term); generalizations of the method 
treating symmetry breaking and bifurcations by uniform approximations; study of
models with the Hamiltonian nonlinear in spin and driven by an external system,
such as the kicked top \cite{haak}; and further applications to specific
systems of interest in spintronics, molecular dynamics, and nuclear physics.

\section*{Acknowledgements}

We are grateful to Matthias Brack for his continuing interest in the project,
many stimulating discussions, and critical reading of the manuscript. Christian
Amann and Metaxi Mehta helped us with the implementation of numerical
computations. Klaus Richter is acknowledged for his helpful comments and
advice. We also would like to thank Jens Bolte, Rainer Glaser, Hermann Grabert,
Stefan Keppeler,  Hajo Leschke, Michael Thoss, and Simone Warzel for the useful
exchange of opinions and critique. This work has been supported by the Deutsche
Forschungsgemeinschaft.

\appendix

\section{Exact quantum-mechanical evaluation of the modulation factor \eq{Ms1}}
\label{ExQM}

\subsection{Exact calculation of the trace of a spin propagator}

We consider a spin in the external magnetic field ${\bf C} (t) \equiv (2 {\rm
Re}\, f (t), - 2 {\rm Im}\, f (t), 2A (t))$ described by the Hamiltonian
\be
  \mathcal H (t) = \hbar\, \hat \mathbf s \cdot \mathbf C (t) = \hbar\, [2A (t)
  \hat s_3 + f (t) \hat s_+ + (f (t))^* \hat s_- ].
\ee
Its exact quantum-mechanical propagator between the spin coherent states $|z_i
\rangle$ and $\langle z_f|$ is \cite{koch}
\be
  K_s (z_f^* , z_i ; T) = \frac{[ a^* (T) - b^* (T)\, z_i + b (T) z^*_f +
  a (T)\, z^*_f z_i]^{2s}}{(1 + |z_f |^2)^s (1 + |z_i |^2)^s}
\label{spex}
\ee
where the coefficients $a (T)$, $b (T)$ are to be found from the equation
\be
  \frac{d}{dt} 
  \left( \begin{array}{cc}
    a & b \\
    -b^* & a^*
  \end{array} \right) 
  = -i \left( \begin{array}{cc}
    A & f \\
    f^* & -A
  \end{array} \right) 
  \left( \begin{array}{cc}
    a & b \\
    -b^* & a^*
  \end{array} \right)
\label{eqsu}
\ee
with the initial conditions $a(0)=1$, $b(0)=0$. One can observe that the
determinant of the matrix
\be
  U(t) = \left( \begin{array}{cc}
    a (t) & b (t) \\
    -b^* (t) & a^* (t)
  \end{array} \right) 
\label{uab}
\ee
remains constant: $|a(t)|^2 +|b(t)|^2 =1$. Hence, $U(t)$ belongs to the group
SU(2) and can be represented in the form
\be
  U(t) = \cos \frac{\eta (t)}{2} + i ({\bf m} (t) \cdot
  \bgrk{\sigma} )\sin \frac{\eta (t)}{2},
\label{um}
\ee
where ${\bf m}^2 =1 $ and $\bgrk{\sigma}$ is a vector of Pauli matrices.
Comparing \eq{uab} and \eq{um} we deduce that
\be
  a(t) = \cos \frac{\eta (t)}{2} + i m_3 (t)\, \sin \frac{\eta (t)}{2} , \quad
  b(t) = i [m_1 (t) - i m_2 (t)] \sin \frac{\eta (t)}{2} . 
\ee
Since
\be
  \frac12 {\bf C} (t) \cdot \bgrk{\sigma} = 
  \left( \begin{array}{cc}
    A (t) & f (t) \\
    f^* (t) & -A (t)
  \end{array} \right),
\ee
we can rewrite \eq{eqsu} as
\be
  \frac{d}{dt} U(t) = - \frac{i}{2} ( {\bf C} (t) \cdot \bgrk{\sigma} )
  U(t) , \quad U(0)=I_2.
\label{dudt}
\ee
Note that this equation has the same form as the equation for a spin-1/2
propagator in the standard $|s,m_s \rangle$ basis. 

The trace of the spin propagator \eq{spex} is
\be
  Z_s (T) = {\rm Tr}\, K_s = \int d \mu_s (z'^*, z')\, K_s (z_f^* = z'^*, z_i
  =z'; T), 
\label{trpr}
\ee
where $d \mu_s (z'^*, z')$ is given by \eq{resun} and
\be
  K_s (z'^* , z' ;T) = \left[ \frac{ a^* (T) - b^* (T) z' + b (T) z'^* +
  a (T) |z'|^2}{1 + |z'|^2} \right]^{2s} .
\label{kzz}
\ee
After the stereographic projection [cf.\ \eq{clsp}]
\be
  n'_1 + i n'_2 \equiv \sin \theta' \exp (i \phi' ) = \frac{2 z'^*}{1 +
  |z'|^2} ,  \quad n'_3 \equiv \cos \theta' = - \frac{1 - |z'|^2}{1 + |z'|^2} 
\ee
and simple transformations we can cast (\ref{kzz}) into the form
\be
  K_s ({\bf n'}; T) = \left( \cos \frac{\eta (T)}{2} + i {\bf n'} \cdot
  {\bf m} (T) \sin \frac{\eta (T)}{2} \right)^{2s}.
\ee
Since the measure
\be
  d \mu_s (z'^*, z') = d \mu_s ({\bf n'}) = \frac{2s+1}{4 \pi} \sin \theta'\, d
  \theta'\, d \phi'
\ee
is invariant under rotations, we can choose the axis $\mathbf e_3$ along ${\bf
m} (T)$. Then
\be
  Z_s (T) = \int d \mu_s ({\bf n'}) K_s ({\bf n'}; T) = \int d \mu_s ({\bf n'})
  \left( \cos \frac{\eta (T)}{2} + i n'_3 \sin \frac{\eta (T)}{2} \right)^{2s}.
\label{intz}
\ee
Making the following transformations in \eq{intz} 
\beq
  Z_s (T) &=& \frac{2s+1}{2} \int_{0}^{\pi}  \left( \cos \frac{\eta (T)}{2} + i
  \cos \theta' \sin \frac{\eta (T)}{2} \right)^{2s} \sin \theta' d \theta'
  \nonumber \\ 
  &=&  \frac{2s+1}{2} \int_{-1}^{1}  \left( \cos \frac{\eta (T)}{2} + i x \sin
  \frac{\eta (T)}{2} \right)^{2s} dx \nonumber \\
  &=& \frac{1}{2i \sin [\eta (T)/2]} \left( \cos \frac{\eta (T)}{2} + i x \sin
  \frac{\eta (T)}{2} \right)^{2s+1} \bigg|^{x=1}_{x=-1} , \nonumber 
\eeq
we obtain the final result
\be
Z_s (T) = \frac{\sin [(s + 1/2)\eta (T)]}{\sin [\eta (T)/2]} ,
\label{zst}
\ee
where $\eta (T)$ is found from $2 \cos [\eta (T)/2] = 2 {\rm Re} [a(T)]  \equiv
{\rm Tr} [U(T)]$.

In order to give the geometric interpretation of $\eta (T)$, consider an
equation for the spin $s$ propagator [cf.\ \eq{dudt}]
\be
  \frac{d}{dt} {\cal D}^s  = -i ({\bf C} \cdot {\bf J}^s) {\cal D}^s ,
  \quad {\cal D}^s (0) = I_{2s+1},
\ee
where ${\bf J}^s = (J_1^s , J_2^s , J_3^s)$ are the generators of the 
$(2s+1)$-dimensional irreducible representation of SU(2). Since ${\cal D}^s (T)
= \exp [i {\bf m} (T) \cdot {\bf J}^s \eta (T)]$, as well as $U(T) = \exp [i
{\bf m} (T) \cdot \bgrk{\sigma} \eta (T)/2]$, $\eta (T)$ has the meaning of the
rotation angle around the ${\bf m} (T)$ axis.

Note that one can alternatively find $Z_s (T)$ as ${\rm Tr} [{\cal D}^s (T)]$.
Choosing ${\bf m} (T)$ as a quantization axis and making simple calculations,
one would obtain the result coinciding with \eq{zst}.

\subsection{Calculation of $\eta (T)$}

For the reader's convenience we derive \eq{stangm} within our notation. We will
closely follow the discussion in \cite{boke}, where this result previously
appeared.

Without loss of generality we can choose the basis in which ${\bf e}_3  = {\bf
m} (T)$. One can decompose $U(t)$ \eq{uab} at every $t$ into the matrix product
\be
  U(t) = 
  \left( \begin{array}{cc}
    \cos (\theta /2) \, e^{-i \phi}& - \sin (\theta / 2)  \\
    \sin (\theta /2) & \cos (\theta /2)\, e^{i \phi}
  \end{array} \right)
  \left( \begin{array}{cc}
    e^{i \psi } & 0 \\
    0& e^{-i \psi }
  \end{array} \right) \equiv U_1 (t) U_0 (t),
\label{decomp}
\ee
where $a = \cos (\theta /2) \, e^{i (\psi - \phi)}$ and $b =  - \sin (\theta /
2)\, e^{-i \psi}$. Decomposition \eq{decomp} corresponds to the choice of a
certain section in the principal U(1) Hopf bundle over $\mathbb S^2$, $\psi$
being a fiber coordinate. Imposing the time periodicity on $U_1(t)$ and
recalling that $U(0)=I_2$, we establish that 
\be
  \psi (T) - \psi(0) = \eta (T) / 2. 
\ee
Note that the decomposition \eq{decomp} is not well defined at $t=0$, i.e.,
when $\theta = 0$. Hence, the initial value $\psi(0)$ is not determined.
(Still, $\eta (T)$ can be found unambiguously.)  Although not essential for our
purposes, the problem can be fixed if we choose a different decomposition near
$\theta = 0$, related to the former by a gauge
transformation~\cite{boke,wuya}. 

We define the projection of the Hopf bundle $(\theta, \phi, \psi) \in \mathbb
S^3 \longrightarrow {\bf n} \in \mathbb S^2$ by
\be
  {\bf n} \cdot \bgrk{\sigma} = U \sigma_3 U^{\dag} = U_1 \sigma_3 U_1^{\dag},
\ee
which is equivalent to $n_1 +i n_2 =\sin \theta \exp (i\phi)$, $n_3 =
\cos\theta$. One can find equations for ${\bf n}(t)$ and $\psi (t)$ from
\eq{dudt}. Thus, one should calculate 
\be
  \dot{{\bf n}} = \frac12 {\rm Tr} \left[ \bgrk{\sigma} \frac{d}{dt} \left( U
  \sigma_3 U^{\dag} \right) \right] 
\ee 
to recover the equation
\be
  \dot{{\bf n}} = {\bf C} \times {\bf n} , \quad {\bf n} (0) = {\bf n} (T).
\label{prec}
\ee
After the stereographic projection \eq{clsp} it becomes
\be
  \dot{z} = - i [ 2 A z +f - f^* z^2 ] , \quad z(0) = z(T).  
\label{zt} 
\ee
To derive an equation for $\psi$ one should calculate 
\be 
  \dot \psi -\frac 1 2 (1 + \cos \theta) \dot \phi = -\frac{i}{2} {\rm Tr}
  \left[ \sigma_3 U^{\dag} \frac{d}{dt} U \right]
\label{ccf}
\ee
taking into account \eq{decomp} and \eq{zt}. In the  language of the fiber
bundle theory, it corresponds to the pull-back of the canonical connection form
[\emph{r.h.s.} of \eq{ccf}] onto the section defined by \eq{decomp}. The result
of the calculation is
\be
\dot{\psi} = \frac{1}{2i} \frac{z \dot{z}^* - z^* \dot{z}}{1 + |z|^2}
- \frac12 {\bf C} \cdot {\bf n} (z^* , z) = [ A - (f^* z + f z^*)/2].
\ee
Integrating it we find
\beq
  \eta (T) &=& 2 [\psi (T) - \psi (0)] = \int_0^T dt \left[\frac{z \dot{z}^* 
  - z^* \dot{z}}{i (1 + |z|^2 )} - {\bf C} \cdot {\bf n} (z^* , z) \right]
  \nonumber \\ 
  &=&  \int_0^T dt \,[2A - (f^* z + f z^*)],
\eeq
where $z(t)$ and $z^* (t)$ are to be found from \eq{zt}.

\section{Adiabatic limit in the extended phase space}
\label{adext}

\begin{figure}[t]
  \vspace*{.5cm}
  \begin{center}
    {\hspace*{0cm}
    \psfig{figure=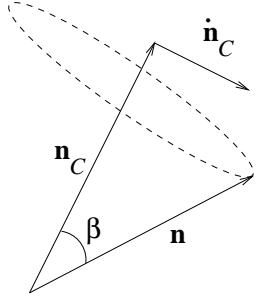,width=0.2 \textwidth}}
  \end{center}
  {\vspace*{0cm}}
  \caption{{\small Classical spin precessing about the instantaneous magnetic
  field.}} 
\label{beta}
{\vspace*{.0 cm}}
\end{figure}

Here we present an intuitive derivation of the trace formula \eq{trfSCL} within
the framework of our semiclassical theory in the extended phase space (Sec.\
\ref{seps}). In the adiabatic regime the classical system, described by the
Hamiltonian $\mathcal H$ \eq{chs}, possesses an adiabatic invariant---the angle
$\beta$ between the instantaneous magnetic field $\mathbf C$ and the classical
spin vector $\mathbf n$. To demonstrate this, let us consider the quantity
$\alpha \equiv \cos \beta = \mathbf n \cdot \mathbf n_C$, where $\mathbf n_C
\equiv \mathbf C/ |\mathbf C|$ (Fig.\ \ref{beta}). Its time derivative averaged
over the fast spin motion is
\be
  \left< \dot \alpha \right> = \left< \mathbf n \cdot \dot {\mathbf n}_C
  \right> \approx 0
\ee
taking into account that $\dot {\mathbf n},\; \dot {\mathbf n}_C \perp \mathbf
n_C$.

Having an approximate integral of motion $\alpha$ has two consequences. First,
the fast spin precession about $\mathbf C$ can be separated from the orbital
motion in the adiabatic limit. The latter then is described by the polarized
Hamiltonian
\be
  \mathcal H_\alpha ({\bf q,p}) = \mathcal H_0 ({\bf q,p}) + \hbar s \alpha
  |{\bf C} ({\bf q}, {\bf p})|
\ee
with $\alpha$ as a parameter. Second, in a system with one integral of motion
in addition to the energy, the periodic orbits are not isolated, but appear in
one-parameter families instead. Hence, we should use the modified trace formula
\cite{crli} that involves the summation over the periodic orbit families
(p.o.f.):
\be
  \delta g_{sc} (E) = - \mathrm{Im}\, \frac 1 {i \pi \hbar}
  \sum_{\mathrm {p.o.f.}} \left[ \frac {2 \pi} {(2 \pi i \hbar)^{1/2}\left|
  \frac {\pd \eta} {\pd J} \right|^{1/2}} \right] \frac {T_{ppo} \exp \left(
  \frac i \hbar {\cal S} + i \varphi^{SK} - i \frac \pi 2 \sigma \right)}
  {\left|\det \left(\widetilde M_{\mathrm{orb}} - I_{2(d-1)} \right)
  \right|^{1/2}},
\label{trfcont}
\ee 
where the factor in the square brackets is absent in the standard Gutzwiller
trace formula for isolated orbits, the stability matrix $\widetilde
M_{\mathrm{orb}}$ is calculated only for the orbital subspace, the action
${\cal S}$, Solari-Kochetov phase $\varphi^{SK}$, and Maslov index $\sigma$ are
evaluated for any representative of a family. $J = \hbar s \alpha$ is the
classical-spin projection on the direction $\mathbf n_C$ of the field and 
\be
  \eta (\alpha) \equiv \eta [\mathbf{q,p}; T_{po}^{(\alpha)}]
\label{etaalpha}
\ee
is the total angle of rotation of the Bloch sphere for a periodic orbit of
$\mathcal H_\alpha$, where the \emph{r.h.s.}\ of \eq{etaalpha} is calculated
according to \eq{etaqpt} (cf.\ Sec.\ \ref{WCL} and Ref.\ \cite{zait}). 
Obviously, we have $\eta \gg 2 \pi$, since the spin vector makes many rotations
during the period of the orbital motion.

\begin{figure}[t]
  \vspace*{.0cm}
  \begin{center}
    {\hspace*{0cm}
    \psfig{figure=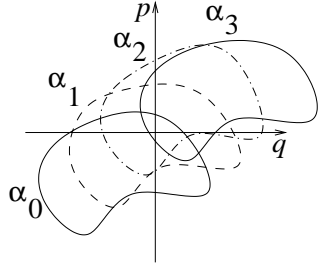,width=0.25 \textwidth}}
  \end{center}
  {\vspace*{0cm}} 
  \caption{{\small Elements of the periodic orbit set in the orbital
  subspace.}} 
{\vspace*{.0 cm}}
\label{pos}
\end{figure}

As a result of the approximate separation of variables, to any classical
trajectory of $\mathcal H_\alpha ({\bf q,p}) = E$ corresponds a family of
classical trajectories in the extended phase space. This family will consist of
periodic orbits if the respective orbit of $\mathcal H_\alpha$ is periodic
\emph{and} the Bloch sphere makes an integer number of rotations during the
orbital period. The latter condition follows from the fact that the whole
family, not just an isolated orbit, is periodic on the Bloch sphere. Thus, we
have a requirement
\be
  \eta (\alpha) = 2 \pi N (\alpha) \quad (N (\alpha) \gg 1)
\label{percond}
\ee
with an integer $N (\alpha)$. Clearly, this condition can be satisfied only for
selected values of $\alpha$. Now we will show that the allowed values of
$\alpha$ form discrete sets that become very dense in the adiabatic limit. 
Consider a periodic orbit of $\mathcal H_{\alpha = -1}$. Let us assume that as
$\alpha$ changes from $-1$ to $1$ for fixed energy, the orbit can be
continuously deformed while maintaining its periodicity and that $\eta
(\alpha)$ increases with $\alpha$. If for $\alpha = -1$ the requirement
\eq{percond} is not fulfilled, we shift $\alpha$ by a small amount to $\alpha =
\alpha_0$, such that $\eta (\alpha_0) = 2 \pi N$. The greater the frequency of
precession $|{\bf C}|$, the smaller the shift. Changing $\alpha$ further to
$\alpha_1 > \alpha_0$, we achieve that the Bloch sphere makes one more rotation
for the newly deformed orbit, i.e., $\eta (\alpha_1) = 2 \pi (N + 1)$. We
continue this process till we reach the upper bound $\alpha_L \leq 1$ for which
$\eta (\alpha_L) = 2 \pi (N + L)$. Thus, we have constructed a set $\{-1 \leq
\alpha_0 < \alpha_1 < \ldots < \alpha_L \leq 1 \}$ of the values of $\alpha$
for which the families of orbits in the extended phase space are periodic. We
will call it the periodic-orbit set (p.o.s.). The better the adiabaticity
condition \eq{adiab} is fulfilled, the larger the number of elements of the
p.o.s.\ $L \gg 1$. The periodic orbits within the set are related by an almost
continuous transformation in the $({\bf q,p})$ space (Fig.\ \ref{pos}). If we
start with a different periodic orbit of $\mathcal H_{\alpha = -1}$, we will
obtain another p.o.s. The sum over the p.o.f.\ in \eq{trfcont} can be
represented as the sum over all sets and the elements within each set, i.e., 
\be
  \sum_{\mathrm {p.o.f.}} \, (\ldots) = \sum_{\mathrm {p.o.s.}}\,
  \sum_{n=N}^{N+L} \, (\ldots) = \sum_{\mathrm {p.o.s.}}\, \sum_{M = -
  \infty}^{\infty} \int_N^{N+L} dn\, e^{2 \pi i M n}\, (\ldots).
\label{resum}
\ee
In the second equality we applied the Poisson summation formula to the inner
sum. 

Our conclusions may seem to disagree with the results of Sec.\ \ref{AL}, where
we have found for the periodic orbits in the extended phase space that $\alpha
\approx \pm 1$. There is no contradiction, however, if we recall that in
\eq{preceq} the magnetic field $\mathbf C$ was predetermined, so that, in
general, the orbital frequency and the overall frequency of precession were
incommensurate. Therefore the spin must have been aligned with the field in
order to have a periodicity in the extended phase space. In the present case,
on the other hand, there is a feedback from the spin to the orbital degrees of
freedom, and for certain initial conditions the frequencies become
commensurate, thus producing the periodic-orbit sets.

\begin{figure}[t]
  \vspace*{.0cm}
  \begin{center}
    {\hspace*{0cm}
    \psfig{figure=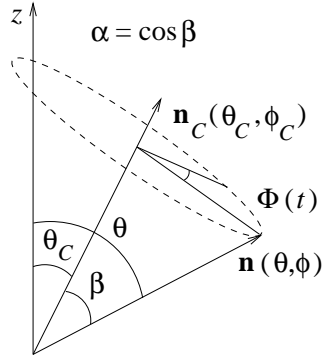,width=0.25 \textwidth}}
  \end{center}
  {\vspace*{0cm}}
  \caption{{\small Illustration of the coordinates' notation. $\Phi
(t) = \int_0^t |{\bf C} ({\bf q}, {\bf p})| dt$}.} 
{\vspace*{.0 cm}}
\label{angles}
\end{figure}

We now turn to the calculation of the action which can be divided into the
orbital and spin parts [cf.\ \eq{perorbact}]. The orbital part $\mathcal
S_{\mathrm{orb}}^{(\alpha)}$ is evaluated in the usual way with the Hamiltonian
$\mathcal H_\alpha ({\bf q,p}) = E$. The spin part is given in terms of the
spin orientation $\mathbf n (\theta, \phi)$ by
\be
  \mathcal S_{\mathrm{spin}}^{(\alpha)} = \hbar s \oint_0^{T_{po}^{(\alpha)}} (
  1 + \cos \theta)\, \dot \phi\, dt = \hbar s [\eta (\alpha) + \alpha \Phi
  (\alpha)] ,
\label{Sspin}
\ee
where $\Phi (\alpha) = \oint_0^{T_{po}^{(\alpha)}} |{\bf C} ({\bf q}, {\bf p})|
dt$. There is a direct analogy between the problem of spin precession and the
problem of a spinning top with fixed base point whose axis of rotation is moved
along a closed circuit. Namely, the magnetic field orientation $\mathbf n_C
(\theta_C, \phi_C)$ corresponds to the axis of rotation, its magnitude $|{\bf
C}|$ becomes the angular velocity of rotation, and the spin orientation
$\mathbf n (\theta, \phi)$ is a direction on the top making angle $\beta$ with
the axis (Fig.\ \ref{angles}). Then the standard theory of adiabatic motion
\cite{hana} provides the relation
\be
  \eta (\alpha) - 4 \pi = \Phi (\alpha) - \Omega (\alpha),
\label{top}
\ee
where $\Omega (\alpha) = \oint_0^{T_{po}^{(\alpha)}} (1 + \cos \theta_C) \,
\dot \phi_C\, dt$ is the solid angle swept over by the axis of rotation in the
lower hemisphere during the period. Using \eq{top} in \eq{Sspin} and taking
into account \eq{percond}, we obtain the spin action
\be
  \mathcal S_{\mathrm{spin}}^{(\alpha)} - 4 \pi \hbar s= 2 \pi \hbar s\,
  (\alpha + 1) N (\alpha)  +  \hbar s \alpha \Omega (\alpha),
\ee
where $N (\alpha)$ was shifted by $-2$.  The second term on the right is the
origin of the Berry phase.

The Solari-Kochetov phase $\varphi^{SK}_\alpha = \eta (\alpha) / 2 = \pi N
(\alpha)$ is equal to an integer multiple of $\pi$. We expect that together
with the Maslov index $\sigma$, it will yield, at the end of the calculation,
the usual Maslov index $\sigma_{po}^{(\alpha)}$ for the trace formula of the
Hamiltonian  $\mathcal H_\alpha ({\bf q,p})$ in the orbital subspace. However,
at this point we have not succeeded in establishing this connection.

We can do the summation over the p.o.f.\ in the trace formula \eq{trfcont} with
the help of \eq{resum}, where the integral can be computed by stationary phase.
For this we express all $\alpha$-dependent quantities in terms of $N (\alpha)$
and then substitute it by a continuous variable $n$. Only $\mathcal
S_{\mathrm{orb}}^{(\alpha)}$ and $2 \pi \hbar s (\alpha + 1) N (\alpha)$
determine the stationary-phase point, the remaining functions are slowly
varying with $n$. The stationary-phase condition is
\be
  0 = 2 \pi M + \frac 1 \hbar \frac \pd {\pd n} \left( \mathcal
  S_{\mathrm{orb}}^{(\alpha)} + 2 \pi \hbar s [\alpha (n) + 1] n \right) =  2
  \pi M + 2 \pi s [\alpha (n) + 1],
\ee
where the result of the classical perturbation theory  
\be
  \frac {\pd \mathcal S_{\mathrm{orb}}^{(\alpha)}} {\pd \alpha} = -
  \oint_0^{T_{po}^{(\alpha)}} \frac {\pd \mathcal H_\alpha}
  {\pd \alpha} dt = - \hbar s \Phi (\alpha) = - 2 \pi \hbar s N (\alpha) +
  \mathcal O (N^0)
\ee
was employed. Thus, the equation for the stationary point $\alpha\,
(n_{\mathrm{st}}) = -M/s -1$ has a solution only if the the \emph{r.h.s.}\ does
not exceed 1 by magnitude. This makes the $M$ sum finite, reducing it to the
sum over quantized polarizations:
\beq
  &~&\sum_{M = - \infty}^{\infty} \int_N^{N+L} dn\, e^{2 \pi i M n + \frac i
  \hbar \left( \mathcal S_{\mathrm{orb}}^{(\alpha)} + 2 \pi \hbar s [\alpha (n)
  + 1] n \right)}\, (\ldots) \nonumber \\
  &~&\approx \sum_{m_s = - s}^s \left\{ e^{\frac i \hbar \mathcal
  S_{\mathrm{orb}}^{(\alpha)}} \,\left[ \frac {(2 \pi i \hbar)^{1/2}\left|
  \frac {\pd \eta} {\pd J} \right|^{1/2}} {2 \pi} \right]\, (\ldots)
  \right\}_{\alpha = m_s/s}.
\label{polsum}
\eeq
$m_s = -M -s$ is either integer or half-integer, depending on the spin $s$. It
labels the quantized spin projection on the $\mathbf n_C$ axis. The prefactor
due to the stationary-phase integration, that appears in the square brackets in
\eq{polsum}, cancels the respective factor in \eq{trfcont}. The outer sum over
the p.o.s.\ in \eq{resum}, if exchanged with the sum over $m_s$, becomes the
sum over the periodic orbits of polarized Hamiltonians ${\cal H}_{\alpha =
m_s/s} \mathbf{(q,p)} \equiv {\cal H}_{\mathrm{eff}}^{(m_s)} \mathbf{(q,p)}$
\eq{Heff}. Thus the density of states is given by the sum of the polarized
trace formulae:
\be
  \delta g_{sc}^{\mathrm{SCL}} (E) = \frac 1 {\pi \hbar} \sum_{m_s =
  - s}^s \sum_{po}\; \frac {T_{ppo}^{(m_s)} \cos
  \left(\frac {S_{po}^{(m_s)}} \hbar  + \varphi_B^{(m_s)} - \frac \pi 2
  \sigma_{po}^{(m_s)} \right)} {\left|\det \left(\widetilde M_{po}^{(m_s)} -
  I_{2(d-1)} \right) \right|^{1/2}}
\ee
with the Berry phase 
\be
  \varphi_B^{(m_s)} = \Bigl. s \alpha\, \Omega (\alpha) \Bigr|_{\alpha =m_s/s}
  = m_s \oint_0^{T_{po}^{(m_s)}} (1 + \cos \theta_C)\, \dot \phi_C\, dt,
\ee
which is equivalent to \eq{Berry}. 

Far from being mathematically rigorous, this derivation illustrates how the sum
over quantized polarizations can be obtained from the classical-spin dynamics,
where any polarization is allowed. As in Sec.\ \ref{AL}, the problem of the
missing no-name term \cite{lifl} remains here. A careful evaluation of the
Maslov indices is also required.

\end{document}